%% file: arXiv_QCE.tex
\documentclass[11pt]{article}
\input{packages.tex}

\input{standard.tex}

\input{AlgMacros}

\usepackage{float}
\newtheorem{theorem}{Theorem}[section]

\newtheorem{definition}{Definition}[section]

\usepackage{tikz-cd}
\newcommand{\thistitle}{Emergent Geometry from Quantum Probability} 
 
\begin{document}

\title{\thistitle}
\author{
Shadi Ali Ahmad$^{a}$ and Marc S. Klinger$^{b}$
	\\ 
	\\
    {\small \emph{\nyu{a}}} \\ \\
	{\small \emph{\uiuc{b}}}
	\\
	}
\date{}
\maketitle
\vspace{-0.5cm}

\begin{abstract}
Carrying the insights of conditional probability to the quantum realm is notoriously difficult due to the non-commutative nature of quantum observables. Nevertheless, conditional expectations on von Neumann algebras have played a significant role in the development of quantum information theory, and especially the study of quantum error correction. In quantum gravity, it has been suggested that conditional expectations may be used to implement the holographic map algebraically, with quantum error correction underlying the emergence of spacetime through the generalized entropy formula. However, the requirements for exact error correction are almost certainly too strong for realistic theories of quantum gravity. In this note, we present a relaxed notion of quantum conditional expectation which implements approximate error correction. We introduce a generalization of Connes’ spatial theory adapted to completely positive maps, and derive a chain rule allowing for the non-commutative factorization of relative modular operators into a marginal and conditional part, constituting a quantum Bayes' law. This allows for an exact quantification of the information gap occurring in the data processing inequality for arbitrary quantum channels. When applied to algebraic inclusions, this also provides an approach to factorizing the entropy of states into a sum of terms which, in the gravitational context, may be interpreted as a generalized entropy. We illustrate that the emergent area operator is fully non-commutative rather than central, except under the conditions of exact error correction. We provide some comments on how this result may be used to construct a fully algebraic quantum extremal surface prescription and to probe the quantum nature of black holes. 
\vspace{0.3cm}
\end{abstract}
\begingroup
\hypersetup{linkcolor=black}
\tableofcontents
\endgroup

\section{Introduction}

At its most fundamental level, quantum theory is a study of non-commutative probability. Instead of working with classical probability distributions and random variables, one is forced by nature to replace these notions with states on Hilbert spaces and non-commuting observables. Despite the drastic differences between classical and quantum probability, they are synthesized in the language of von Neumann algebras~\cite{Takesaki1, Takesaki2, Takesaki3} with Abelian algebras encoding ordinary measure spaces, while non-Abelian ones encode what are often referred to as non-commutative measure spaces. In both its commutative and non-commutative avatars, the study of probability leads naturally to questions in information theory~\cite{Har28, shannon, kolmogorov65, Witten_2020}.
The answers to these questions invariably center around the computation of measures of entropy. Entropic measures have been found to be universal across a wide array of different fields, which is a testament to the ubiquity of information theory. 

A seminal result in classical probability theory is Bayes' law~\cite{Bayes:1764vd}: the joint probability distribution of two random variables can be decomposed into the product of a marginal and a conditional probability distribution. Measures like the entropy and the relative entropy inherit this decomposition and factorize into sums of marginal and conditional contributions. The act of reducing a total distribution into these components relies upon the ability to `integrate out' one of the variables, akin to the partial trace of quantum mechanics. More generally, this operation is described by a classical conditional expectation. On the other hand, given a von Neumann algebra that encodes the quantum observables of two distinct subsystems, we are not guaranteed any factorization due to entanglement. Moreover, there are several distinct quantum operations that resemble classical conditional expectations. These considerations severely complicate the problem of promoting Bayes' theorem and the enterprise of conditional probability to the quantum realm. 

Notwithstanding these challenges, the main contribution of this note is to derive a version of Bayes' law that is valid for \textit{any} von Neumann algebra independent of its type classification or factorizability. This allows us to decompose (relative) density operators, which are algebraic analogues of probability distributions, into a non-commutative product of marginal and conditional operators. In turn, this factorization gives rise to a quantum notion of marginal and conditional entropy which enhances the computation and interpretation of various quantum information theoretic quantities. The non-commutativity of the underlying quantum probability theory manifests itself in a series of new terms appearing in the quantum conditional (relative) entropy which signal the entanglement between subsystems.   

A principal application of our non-commutative Bayes' theorem quantifies the relative entropy between states under the application of a completely positive map. Such maps are ubiquitous in quantum information because they preserve the probabilistic interpretation of quantum theory; for example, measurement can be modeled by such `quantum operations'~\cite{nielsen00}. The relative entropy factorizes as a sum of the relative entropy between states before the application of the quantum operation and a leftover term which \emph{defines} the quantum conditional relative entropy. A cornerstone of quantum information theory is the data processing inequality, which tells us that the relative entropy between states must decrease under the application of a quantum operation~\cite{Cover2006,beaudryrenner}.\footnote{In other words, quantum operations serve to make states less distinguishable.} The quantum conditional relative entropy therefore provides an exact quantification of the information which is lost under the application of the quantum operation. This is in direct analogy with the classical conditional relative entropy which quantifies information loss under the application of a classical channel. In the context of quantum error correction, the conditional relative entropy quantifies obstructions to the invertibility of a noisy quantum channel~\cite{steane, sho95, bennet, em, kf}.

One of the areas which has benefited enormously from the incorporation of quantum information theoretic tools is the study of quantum gravity. Among the central problems in this field is understanding the microscopic origin of entropy and entanglement and its relationship to the emergence of spacetime geometry. Hawking radiation~\cite{hawking1975particle} provided a short-lived puzzle regarding the second law of thermodynamics applied to black holes. Given that black holes both (1) have an entropy proportional to their area~\cite{bekenstein1973black} and (2) radiate and shrink away, it seems that the entropy is allowed to decrease. However, this points to an enhancement of the entropy by a term which measures the entropy of the radiated particles themselves. The quantity combining both the area and the radiation entropy is called the generalized entropy~\cite{bekenstein1974generalized}, and satisfies the second law of thermodynamics~\cite{bardeen1973four}. The generalized entropy may be interpreted as a harbinger of holography. In the AdS/CFT correspondence, bulk physics is \textit{encoded} in a non-local way in a boundary theory. The problem of building geometry from the data of a holographic CFT amounts to identifying boundary observables which are dual to geometric quantities. The seminal work of Ryu-Takayanagi \cite{Ryu:2006bv} identified the entropy of the vacuum on a CFT subregion with the generalized entropy of a particular bulk subregion including the contribution of an extremal area term. 


In \cite{harlow_ryutakayanagi_2017}, it was recognized that such a decomposition of the entropy is reminiscent of exact quantum error correction. This observation was given an algebraic treatment in \cite{faulkner2020holographic}, where it was shown that exact complementary error correction implies the existence of a conditional expectation\footnote{This is a strict form of quantum conditional expectation, as we will discuss in the body of the note.} mapping from the boundary to the bulk which implements the holographic map algebraically. The appearance of the area is a generic feature of the factorization of states in the code-subspace of such a conditional expectation. Unfortunately, these results were largely derived under stringent conditions that do not immediately apply to quantum field theory and gravity. Namely, the algebras involved are typically assumed to be of type I, and the existence of conditional expectations are taken as a given. A byproduct of these strong assumptions is that the emergent area operator is pushed into the center of the bulk algebra. As we shall address, the centrality of the area operator is an indication that important quantum gravitational effects are being suppressed. 

On the other hand, exciting developments in the algebraic approach to quantum field theory and gravity have presented an alternative, purely Lorentzian, derivation of the generalized entropy by means of a mathematical operation called the modular crossed product~\cite{Witten:2021unn, Chandrasekaran:2022cip,Chandrasekaran:2022eqq, Jensen:2023yxy, AliAhmad:2023etg, Klinger:2023tgi, Kudler-Flam:2023hkl, ahmad2024semifinitevonneumannalgebras}. One remarkable feature of the modular crossed product is that it regulates familiar divergences in the von Neumann entropy of quantum field theories and gravity by appending a new operator to a QFT subregion algebra which plays the role of a regulated area. One might hope that this approach to the generalized entropy is compatible with the quantum error correcting perspective addressed above. Unfortunately, the modular crossed product utilized in this setting does not admit a conditional expectation to the original algebra that was used to construct it. Thus, if there is error correction in this case it is at best approximate. 

This brings us to the second main application of our non-commutative Bayes' theorem; it provides an approach to the generalized entropy which bridges the gap between the two points of view here described. Given an inclusion of von Neumann algebras $i: N \hookrightarrow M$ in which $M$ is assumed to be semi-finite, we can use our formalism to express the density operator of a state on $M$ as a non-commutative product of its associated marginal and conditional densities. The entropy of such a state can immediately be factorized into a sum of the `entropy' of the marginal state\footnote{This entropy will require careful interpretation if $N$ is a type III algebra.} and the expectation value of an operator which, in gravitational contexts, we interpret as the area. In this derivation, the area is a fully quantum object. In fact, we will show that the area is \textit{not} central unless the state in question belongs to the code-subspace of a conditional expectation. In other words, the area operator becomes central only when exact quantum error correction is allowed. Using this approach, we can reproduce the entropy computations of \cite{harlow_ryutakayanagi_2017,faulkner2020holographic} under the assumption of exact complementary recoverability, and the entropy computations of \cite{Witten:2021unn, Chandrasekaran:2022cip,Chandrasekaran:2022eqq, Jensen:2023yxy, AliAhmad:2023etg, Klinger:2023tgi, Kudler-Flam:2023hkl, ahmad2024semifinitevonneumannalgebras} when the algebra $M$ is the modular crossed product of a type III algebra.\footnote{In fact, this approach allows for the generalized entropy to be derived for a more broad set of states than those which are considered in \cite{Witten:2021unn, Chandrasekaran:2022cip,Chandrasekaran:2022eqq, Jensen:2023yxy, AliAhmad:2023etg, Klinger:2023tgi, Kudler-Flam:2023hkl, ahmad2024semifinitevonneumannalgebras}.} In both cases the expectation value of the area operator computes the quantum conditional entropy, thereby unifying the information theoretic nature of the geometric contribution to the generalized entropy. What's more, additional contributions to the quantum conditional entropy beyond the classical/exact error correcting case capture quantum gravitational backreaction.

The structure of this paper is as follows. In Section~\ref{sec: classicalrevi}, we review the classical version of Bayes' law and  demonstrate how it leads to a factorization of both the entropy and the relative entropy. We introduce the classical data processing inequality, discuss its role in defining sufficient statistics, and identify the classical relative conditional entropy as quantifying the information gap. This section provides a benchmark for the remainder of the note which seeks to upgrade its insights to the context of general von Neumann algebras. 

In Section \ref{sec: CP Maps} we review the necessary mathematical background for deriving our main result. First, we introduce completely positive maps and discuss different special cases which can be understood as quantum notions of conditional expectations including strict conditional expectations and generalized conditional expectations. In Section \ref{sec: GCEs} we recall the definition of the Petz dual of a quantum channel, and identify generalized conditional expectations as Petz duals of inclusion maps. This provides a mechanism for factorizing states on arbitrary von Neumann algebras into the composition of a marginal state and a quantum channel. In Section \ref{sec: Error Correction} we provide an overview of the interplay between Petz dual maps and quantum error correction. We establish that a generalized conditional expectation becomes a strict conditional expectation if and only if it satisfies a modularity property that predicates a large degree of tensor factorizability. This is closely related to exact quantum error correction. By contrast, we regard generalized conditional expectations as giving rise to only partial factorizability of algebras, and by extension approximate quantum error correction. We present Petz's criterion for the sufficiency of a quantum channel with respect to a family of states. This is naturally related to a saturation of the data processing inequality. Finally, in Section \ref{sec: Spatial Theory} we review Connes' spatial theory for states on von Neumann algebras which is an enhancement of Tomita-Takesaki theory. We describe a generalization of spatial theory which allows us to define the spatial derivative of one completely positive map with respect to another. 

In Section \ref{sec: DPI}, we derive our main result. Our novel contribution is to develop a chain rule for spatial derivatives which relates the spatial derivative between compositions of completely positive maps to products of the spatial derivatives of the maps being composed. In the special case where we regard states on an algebra $M$ as a composition of a marginal state on $N$ and a quantum channel from $M$ to $N$, our chain rule allows us to write down a formula for the relative density operator for states on $M$ in terms of the relative density operator of their marginal states on $N$ acted upon adjointly by an operator we define as the relative conditional density. This is the non-commutative form of Bayes' law. In Section \ref{sec: applications}, we use our non-commutative Bayes' law to show that the area term appearing in the generalized entropy formula can be understood as a quantum conditional entropy in both the exact quantum error correcting regime and the modular crossed product for subregion algebras. Building upon this point, we provide an outline as to how our work might be used to enrich our understanding of holographic subregion duality and black hole physics beyond the semiclassical limit. We conclude in Section \ref{sec: Discussion} in which we summarize and contextualize our results, and provide suggestions for future work. 

\section{Factorization of classical entropy} \label{sec: classicalrevi}
In this section, we give a rapid review of the classical form of Bayes' theorem \cite{Bayes:1764vd}. Given a pair of measure spaces we can always decompose probability distributions on their Cartesian product into the product of a marginal and a conditional probability distribution. The factorization of such a probability distribution naturally implies a factorization of its entropy, which resembles the generalized entropy. 

Consider a measure space which is a Cartesian product $X = X_1 \times X_2$. A probability distribution on $X$ is a map $p: X \rightarrow \mathbb{R}_+$ which is normalized:
\beq
	\int_{X} dx_1 \; dx_2 \; p(x_1,x_2) = 1,
\eeq
Bayes' theorem tells us that $p(x_1,x_2)$ can be factorized into products of the form
\beq \label{Classical Factorization}
	p(x_1,x_2) = p_1(x_1) p_{2 \mid 1}(x_2 \mid x_1) = p_{2}(x_2) p_{1 \mid 2}(x_1 \mid x_2)
\eeq
where
\beq
	p_{1}(x_1) \equiv \int_{X_2} dx_2 \; p(x_1,x_2), \qquad p_{2}(x_2) \equiv \int_{X_1} dx_1 \; p(x_1,x_2),
\eeq
are the \textit{marginal} distributions obtained from $p(x_1,x_2)$. The remaining factors
\beq
	p_{2 \mid 1}(x_2 \mid x_1) \equiv p(x_1,x_2) p_{1}(x_1)^{-1}, \qquad p_{1 \mid 2}(x_1 \mid x_2) \equiv p(x_1,x_2) p_{2}(x_2)^{-1} \label{bayes 1}
\eeq
are called the \textit{conditional} probability distributions for $X_2 (X_{1})$ given $X_1 (X_{2})$. By definition, $p_{2 \mid 1}(x_2 \mid x_1)$ is a probability distribution on $X_2$ for each fixed value of $x_1$ (and similarly for $p_{1 \mid 2}(x_1 \mid x_2)$). 

The product factorization \eqref{Classical Factorization} allows us to factorize the entropy of the distribution $p(x_1,x_2)$ as a sum
\begin{flalign} \label{Classical Generalized Entropy}
	S(p) &= - \int_{X} dx_1 \; dx_2 \; p(x_1,x_2) \ln p(x_1,x_2) \nonumber \\
	&= - \int_{X_1} dx_1 \; p_{1}(x_1) \ln p_{1}(x_1) - \int_{X_1} dx_1 \; p_{1}(x_1) \int_{X_2} dx_2 \; p_{2 \mid 1}(x_2 \mid x_1) \ln p_{2 \mid 1}(x_2 \mid x_1) \nonumber \\
	&= S(p_{1}) + \mathbb{E}_{p_1}\bigg(S(p_{2 \mid 1})\rvert_{X_1}\bigg),
\end{flalign}
where we have focused on the factorization of $p(x_{1}, x_{2}) = p_{1}(x_{1}) p_{2|1}(x_{2}|x_{1})$ without loss of generality.  Here, $S(p_{1})$ is the entropy of the marginal distribution on $X_1$ and $\mathbb{E}_{p_1}\bigg(S(p_{2 \mid 1})\rvert_{X_1}\bigg)$ is the expectation value of the function on $X_1$ given by
\beq \label{Classical Conditional Entropy}
	S(p_{2 \mid 1})\rvert_{x_1} = -\int_{X_2} dx_2 \; p_{2 \mid 1}(x_2 \mid x_1) \ln p_{2 \mid 1}(x_2 \mid x_1).
\eeq
The quantity above is called the conditional entropy, as it computes the entropy of the probability distribution over $X_2$ which is realized for each value $x_1 \in X_1$. Eqn.~\eqref{Classical Generalized Entropy} therefore reads that the entropy of the joint distribution $p(x_1,x_2)$ is the entropy of the marginal $p_{1}(x_1)$ plus the expected entropy of the conditional $p_{2 \mid 1}(x_2 \mid x_1)$ with respect to the marginal. 

Given a pair of distributions $p(x_1,x_2) = p_1(x_1) p_{2 \mid 1}(x_2 \mid x_1)$ and $q(x_1,x_2) = q_1(x_1) q_{2 \mid 1}(x_2 \mid x_1)$ we can also compute the relative entropy. The relative entropy possesses a similar factorization as
\begin{flalign} \label{Classical Data Processing 1}
	D(p \parallel q) &\equiv \int_{X} dx_1 \; dx_2 \; p(x_1,x_2) \ln \frac{p(x_1,x_2)}{q(x_1,x_2)} \nonumber \\
	&= \int_{X_1} dx_1 \; p_1(x_1) \ln \frac{p_1(x_1)}{q_1(x_1)} + \int_{X_1} dx_1 \; p_1(x_1) \int_{X_2} dx_2 \; p_{2 \mid 1}(x_2 \mid x_1) \ln \frac{p_{2 \mid 1}(x_2 \mid x_1)}{q_{2 \mid 1}(x_2 \mid x_1)} \nonumber \\
	&= D(p_1 \parallel q_1) + \mathbb{E}_{p_1}\bigg(D(p_{2 \mid 1} \parallel q_{2 \mid 1})\rvert_{X_1}\bigg). 
\end{flalign}
Here, $D(p_1 \parallel q_1)$ is the relative entropy between the marginalizations of $p(x_1,x_2)$ and $q(x_1,x_2)$ to $X_1$ and $\mathbb{E}_{p_1}\bigg(D(p_{2 \mid 1} \parallel q_{2 \mid 1})\rvert_{X_1}\bigg)$ is the expectation value of the function on $X_1$ given by
\beq
	D(p_{2 \mid 1} \parallel q_{2 \mid 1})\rvert_{x_1} = \int_{X_2} dx_2 \; p_{2 \mid 1}(x_2 \mid x_1) \ln \frac{p_{2 \mid 1}(x_2 \mid x_1)}{q_{2 \mid 1}(x_2 \mid x_1)}. 
\eeq
Rearranging Eqn.~\eqref{Classical Data Processing 1} we obtain
\beq \label{Classical Data Processing 2}
	D(p \parallel q) - D(p_1 \parallel q_1) = \mathbb{E}_{p_1}\bigg(D(p_{2 \mid 1} \parallel q_{2 \mid 1})\rvert_{X_1}\bigg).
\eeq
The left hand side of Eqn.~\eqref{Classical Data Processing 2} is the difference in the distinguishability of the distributions $p$ and $q$ before and after marginalization. 

We refer to the right hand side of Eqn.~\eqref{Classical Data Processing 2} as the conditional relative entropy. If the conditional relative entropy vanishes, we say that the subspace $X_1$ is sufficient with respect to the distributions $p$ and $q$. This means that all of the information in $X$ which can be used to differentiate between $p$ and $q$ is, in fact, contained in $X_1$. One can interpret Eqn.~\eqref{Classical Data Processing 2} as a classical analogue of the data processing inequality since the relative entropy is non-negative. 

In large part, the contribution of the present note is to generalize the insights of Eqs.~\eqref{Classical Factorization}, \eqref{Classical Generalized Entropy}, and \eqref{Classical Data Processing 2} to the fully quantum context. Carrying the intuition of this section to quantum theory is complicated by at least three subtle points:
\begin{enumerate}
	\item There are several notions of conditional expectations for von Neumann algebras. In the commutative case, these notions all coincide, but the same is not true for general non-commutative algebras. The conditional expectation in the classical case is given by marginalization and in the Type I case by the partial trace. 
	\item Quantum systems need not, and typically do not, possess neat factorizations analogous to the Cartesian product structure assumed in the above analysis. This is an expression of quantum entanglement.
	\item We are not guaranteed to possess density operator representations of states in non-commutative operator algebras.
\end{enumerate}
In certain special cases these problems can be overcome and a quantum analog of \eqref{Classical Generalized Entropy} is obtained. A significant example of this is the case of exact quantum error correction for semi-finite von Neumann algebras \cite{faulkner2020holographic}. However, exact quantum error correction overcomes the above three concerns by (1) relying upon the existence of a particularly strong notion of quantum conditional expectation (2) ensuring a factorization of the algebras involved, and, at least in most of the literature, (3) restricting attention to type I von Neumann algebras so that problems of semi-finiteness are avoided.\footnote{There has been some work aimed towards understanding entanglement wedge reconstruction in the context of approximate error correction, however these typically assume that the algebras involved are finite dimensional. See e.g. \cite{Cotler:2017erl}.}

In this note we introduce a new approach to generalizing \eqref{Classical Factorization}, \eqref{Classical Generalized Entropy}, and \eqref{Classical Data Processing 2} to the quantum context which envelops exact quantum error correction as a special case. One of the main observations of this analysis is that in instances in which exact error correction is not possible, it is necessary to replace standard conditional expectations with more relaxed objects called completely positive maps. Even without a standard conditional expectation, states can still be factorized in an appropriate manner provided there exists an inclusion of operator algebras. Using a generalization of spatial theory adapted to completely positive maps, factorized states can be made to give rise to factorized relative density operators that are well defined even in the context of type III von Neumann algebras, providing a quantum version of \eqref{Classical Factorization}. This allows for an analog of \eqref{Classical Data Processing 2} to be obtained at the level of the relative entropy which reduces to an entropy factorization of the form \eqref{Classical Generalized Entropy} in the semi-finite case. In the holographic context the quantum conditional entropy allows us to define a `relative area operator' which is generically non-central, thereby giving rise to a series of quantum corrections to the existing results. From a more general point of view, this result can be interpreted as quantifying violations to the saturation of the data processing inequality which governs the sufficiency of quantum channels with respect to families of states.

\section{Completely positive maps and non-commutative conditional expectations} \label{sec: CP Maps}
In this section, we review essential elements of the theory of completely positive maps in the context of von Neumann algebras. Our aim is to extend the results of the previous section to the non-commutative realm, which are underpinned by the uniqueness of the conditional expectation of probability theory in Eqn.~\eqref{Conditional Expectation}. In the quantum setting, there are three separate notions of such maps. We show that for any inclusion of von Neumann algebras, there exists at least one of these. This allows us to make a connection with quantum error correction and the Petz map. Finally, since von Neumann algebras are represented on Hilbert space, we review the \textit{spatial theory} associated to completely positive maps, which is a generalized approach to modular theory.

\subsection{Motivation and definitions}
In quantum information theory, the set of allowable quantum operations and channels one applies to systems is populated by completely positive maps. The basic idea behind such maps is ensuring that density operators which house the probabilities of measurements for the \textit{total} system remain valid density operators under operations on any subsystem. A generic operation on a subsystem which is entangled with its complement may introduce, due to entanglement, negative eigenvalues of the resultant density operator thus breaking the probabilistic interpretation of quantum mechanics. Thus, one needs to restrict the set of allowable operations to those that preserve positivity of probabilities after adjoining \textit{any} subsystem to a given one. 

The definition of a completely positive map can be formulated at the level of $C^{*}$ algebras. 

\begin{definition}[Completely positive]
    Let $A$ and $B$ be a pair of $C^*$ algebras and $\alpha: A \rightarrow B$ a linear map. The map $\alpha$ is called $n$-positive for $n \in \mathbb{N}$ if the extension $\alpha_n: A \otimes \mathbb{C}^{n \times n} \rightarrow B \otimes \mathbb{C}^{n \times n}$ given by $\alpha_n(a \otimes m) \equiv \alpha(a) \otimes m$ is positive. If $\alpha_n$ is positive for each $n \in \mathbb{N}$, then $\alpha$ is called completely positive. 
\end{definition}
Now, let $\alpha$ be a completely positive map\footnote{In fact, we will allow $\alpha$ to map into the set of operators affiliated with $B(H)$ so that densely defined but unbounded operators may be included in our analysis.} $\alpha: A \rightarrow B(H)$ where $H$ is some Hilbert space. Stinespring's dilation theorem \cite{stinespring1955,CHOI1975285} tells us that there will always exist a representation $\pi_{\alpha}: A \rightarrow B(H_{\alpha})$ along with a map $W_{\alpha \mid H}: H_{\alpha} \rightarrow H$ such that
\beq \label{Stinespring}
	\alpha(a) = W_{\alpha \mid H}^{\dagger} \pi_{\alpha}(a) W_{\alpha \mid H}. 
\eeq
Any representation space $H_{\alpha}$ for which Eqn.~\eqref{Stinespring} holds is called spatial for $\alpha$. 

For our purposes, it is useful to further classify the space of completely positive maps in order to define three separate candidates for non-commutative conditional expectations. Again, let $\alpha: A \rightarrow B(H)$ be a completely positive map, and suppose that $\pi_B: B \rightarrow B(H)$ is a representation of $B \subset A$. We denote the inclusion of $B$ into $A$ by $i: B \hookrightarrow A$. There are three properties we can require of $\alpha$ in addition to complete positivity. They are:
\begin{enumerate}
    \item 	\textbf{Unitality:} Provided $A$ is a unital algebra, the map $\alpha$ is called unital if 
	\beq
		\alpha(\mathbb{1}_A) = \mathbb{1}_H.
	\eeq 
 If $\alpha$ is spatial on $H_{\alpha}$ and unital, the map $W_{\alpha \mid H}$ which implements $\alpha$ will be an isometry which can be seen from
\begin{equation}
     \alpha(\mathbb{1}_{A} ) = W^{\dagger}_{\alpha \mid H} W_{\alpha \mid H} = \mathbb{1}_{H}.\nonumber
     \end{equation}
\item \textbf{$B$-homogeneity:} The map $\alpha$ is called $B$-homogeneous if for each $a \in A$ and $b_1,b_2 \in B$ we have
	\beq \label{B-homogeneity}
		\alpha(i(b_1) \; a \; i(b_2)) = \pi_B(b_1) \alpha(a) \pi_B(b_2).
	\eeq
	If $\alpha$ is spatial on $H_{\alpha}$ and $B$-homogeneous, the map $W_{\alpha \mid H}$ which implements $\alpha$ is a $B$ intertwiner
	\beq
		W_{\alpha \mid H} \pi_B(b) = \pi_{\alpha} \circ i(b) W_{\alpha \mid H},  \nonumber
	\eeq
 which can be seen from plugging the spatial condition in Eqn.~\eqref{Stinespring} into Eqn.~\eqref{B-homogeneity}. 
 \item \textbf{$B$-preservation:} The map $\alpha$ is called $B$-preserving if
	\beq
		\alpha(A) \subset \pi_B(B) \subset B(H). 
	\eeq
	In this case we can enhance $\alpha$ to a map between the algebras, $\alpha_B: A \rightarrow B$ given by $\alpha_B \equiv \pi_B^{-1} \circ \alpha$. 

\end{enumerate}

A completely positive map which is unital is called a \emph{quantum channel}. A completely positive map which is unital and $B$-preserving is called a \emph{generalized conditional expectation}.\footnote{This is when we assume that $B \subset A$. More generally, if $\alpha$ is unital and $B$-preserving for an algebra $B$ not assumed to be contained in $A$, then $\alpha_B \equiv \pi_B^{-1} \circ \alpha: A \rightarrow B$ is a quantum channel between unital $C^*$ algebras.} A completely positive map which is $B$-homogeneous and $B$-preserving is called an \emph{operator valued weight}.\footnote{It is worth noting that an operator valued weight gives $A$ the structure of a $B$-module with $B$-valued inner product $G_{\alpha}(a_1,a_2) \equiv \alpha(a_1^* a_2)$. This property will allow such an object to be used to define algebra extensions, which is the topic of forthcoming work \cite{ali-ahmad2024_QSystem}.} Finally, a completely positive map which is unital, $B$-homogeneous, and $B$-preserving is what is conventionally referred to as a \emph{conditional expectation} in the non-commutative context.\footnote{One may also wonder whether any signficance can be given to completely positive maps which are unital and $B$-homogeneous. In the literature, these are called \emph{conditional states}, and have been used to study the conditional entropy from a different point of view.}

\subsection{Petz duality and generalized conditional expectations} \label{sec: GCEs}
It can be shown that, given any inclusion of von Neumann algebras $i: N \hookrightarrow M$, one can always construct a generalized conditional expectation. This result follows from Petz's duality theory for maps between von Neumann algebras with faithful, semifinite, normal weights. The presentation of this section is drawn from various of Petz's works including \cite{PetzDuality,Petz:1988usv,qentropy}.  


Consider a pair $(M,\varphi)$ with $M$ a von Neumann algebra and $\varphi \in P(M)$ a faithful, semifinite, normal weight. Such an object also implicates a GNS Hilbert space characterized by the triple $(L^2(M;\varphi), \eta_{\varphi},\pi_{\varphi})$ where $L^2(M;\varphi)$ is the Hilbert space obtained by closing $\mathfrak{n}_{\varphi} \equiv \{x \in M \; | \; \varphi(x^*x) < \infty\}$ in the inner product\footnote{Here, $\eta_{\varphi}: \mathfrak{n}_{\varphi} \rightarrow \mathfrak{n}_{\varphi}/\mathfrak{k}_{\varphi}$ is a projection, where $\mathfrak{k}_{\varphi} \equiv \{x \in M \; | \; \varphi(x^* x) = 0\}$.}
\beq \label{GNS Inner Product}
	g_{L^2(M;\varphi)}(\eta_{\varphi}(x),\eta_{\varphi}(y)) \equiv \varphi(x^*y), \; x,y \in \mathfrak{n}_{\varphi}. 
\eeq
The map $\pi_{\varphi}: M \rightarrow B(L^2(M;\varphi))$ is a standard representation with $\pi_{\varphi}(x) \bigg(\eta_{\varphi}(y)\bigg) = \eta_{\varphi}(xy)$. 

In addition to the inner product Eqn.~\eqref{GNS Inner Product}, we can also define the KMS inner product on $M$:
\beq
	g^{KMS}_{\varphi}(x,y) \equiv g_{L^2(M;\varphi)}\bigg(\eta_{\varphi}(x), J_{\varphi} \eta_{\varphi}(y)\bigg) = g_{\varphi}\bigg(\eta_{\varphi}(x^*), \Delta_{\varphi}^{1/2} \eta_{\varphi}(y)\bigg).
\eeq
Here, $S_{\varphi} = J_{\varphi} \Delta_{\varphi}^{\frac{1}{2}}$ is the polar decomposition of the Tomita operator of modular theory. The map $J_{\varphi}$ is the modular conjugation and $\Delta_{\varphi}$ is the modular operator of $\varphi$. Given two pairs $(M_1,\varphi_1)$ and $(M_2,\varphi_2)$ and a quantum channel $\alpha: M_1 \rightarrow M_2$, there exists a unique conjugate map $\alpha_{\varphi_1,\varphi_2}^{\dagger}: M_2 \rightarrow M_1$ which serves as the formal adjoint of $\alpha$ in the KMS inner products of $M_{1}$ and $M_{2}$, namely
\beq
	g^{KMS}_{\varphi_1}(m_1, \alpha_{\varphi_1,\varphi_2}^{\dagger}(m_2)) = g^{KMS}_{\varphi_2}(\alpha(m_1),m_2), \; \forall m_1 \in M_1, m_2 \in M_2. 
\eeq	
 In this respect, $\alpha_{\varphi_1,\varphi_2}^{\dagger}$ depends upon the choices of the states $\varphi_1$ and $\varphi_2$. The map $\alpha_{\varphi_1,\varphi_2}^{\dagger}$ is called the Petz dual of $\alpha$ with respect to the states $\varphi_1$ and $\varphi_2$.

 

Petz proved that if $\alpha$ is completely positive and unital, then so too is $\alpha_{\varphi_1,\varphi_2}^{\dagger}$. Thus, the Petz dual of a quantum channel is itself a quantum channel. The Petz dual is always spatially implemented on the Hilbert space $L^2(M_2;\varphi_2)$. To define its spatial implementation, we first introduce the map
\beq
    V_{\alpha}: L^2(M_1;\varphi_1) \rightarrow L^2(M_2;\varphi_2), \qquad V_{\alpha}(\eta_{\varphi_1}(m_1)) = \eta_{\varphi_2} \circ \alpha(m_1). 
\eeq
which implements the channel $\alpha$ at the Hilbert space level. Then,
\beq
	\alpha_{\varphi_1,\varphi_2}^{\dagger} = \pi_{\varphi_1}^{-1} \circ \text{Ad}_{W^{\dagger}} \circ \pi_{\varphi_2},
\eeq
with $W \equiv J_{\varphi_2} V_{\alpha} J_{\varphi_1}$. That is
\beq \label{Petz Spatial}
	\pi_{\varphi_1} \circ \alpha_{\varphi_1,\varphi_2}^{\dagger}(m_2) = J_{\varphi_1} V_{\alpha}^{\dagger} J_{\varphi_2} \pi_{\varphi_2}(m_2) J_{\varphi_2} V_{\alpha} J_{\varphi_1}. 
\eeq
Eqn.~\eqref{Petz Spatial} is the form of the Petz map which typically appears in literature on quantum error correction~\cite{qentropy,faulkner2020approximaterecoveryrelativeentropy}.

Provided $N \subset M$ is a von Neumann inclusion and $\varphi \in P(M)$ is a weight with $\varphi_0 \equiv \varphi \circ i \in P(N)$, the inclusion $i: N \hookrightarrow M$ can be regarded as a quantum channel.\footnote{We shall always assume that our inclusions are unital.} In this case $L^2(N;\varphi_0) \subset L^2(M;\varphi)$ is a Hilbert subspace and the adjoint of the contraction $V_{i}^{\dagger}: L^2(M;\varphi) \rightarrow L^2(N;\varphi_0)$ can be taken to be an orthogonal projection. The Petz dual
\beq
	i^{\dagger}_{\varphi} \equiv i_{\varphi_0,\varphi}^{\dagger} = \pi_{\varphi_0}^{-1} \circ \text{Ad}_{J_{\varphi_0} V^{\dagger}_i J_{\varphi}} \circ \pi_{\varphi}
\eeq	
is a unital $N$-preserving map and thus a generalized conditional expectation. This map was originally introduced by Accardi and Cecchini \cite{ACCARDI1982245}. As they observed, it has the useful property of factorizing the state as $\varphi = \varphi_0 \circ i_{\varphi}^{\dagger}$.

\subsection{Quantum Sufficiency of Channels} \label{sec: Error Correction}

As we have addressed, a generalized conditional expectation is both unital and subalgebra preserving. However, it is generally not homogeneous. It turns out that the context in which the generalized conditional expectation is moreover equivalent to a bona-fide conditional expectation is intimately related with the problem of quantum error correction, which we now review. 

Let $\alpha: N \rightarrow M$ be a quantum channel and $\varphi \in P(M)$ a weight such that $\varphi \circ \alpha \in P(N)$. We will refer to $\varphi$ as the input state and $\varphi \circ \alpha$ as the output. The Petz dual $\alpha_{\varphi}^{\dagger} \equiv \alpha_{\varphi \circ \alpha,\varphi}^{\dagger}: M \rightarrow N$ is itself a channel, and $\varphi = \varphi \circ \alpha \circ \alpha_{\varphi}^{\dagger}$. We begin our discussion with a central result:
\begin{theorem}[Recoverability]
For any $n \in N$
\beq
	\alpha^{\dagger}_{\varphi} \circ \alpha(n) = n \iff \alpha(n^*n) = \alpha(n)^* \alpha(n), \; \alpha \circ \sigma^{\varphi \circ \alpha}_t(n) = \sigma^{\varphi}_t \circ \alpha(n) \; \forall t \in \mathbb{R}. 
\eeq
\end{theorem} 
One can think of this theorem as identifying the set of elements $n \in N$ which can be recovered by the Petz dual after the implementation of the channel $\alpha$:
\beq
	N_{\alpha,\varphi} \equiv \{n \in N \; | \; \alpha^{\dagger}_{\varphi} \circ \alpha(n) = n\}. 
\eeq

In the case that $i: N \hookrightarrow M$ is an inclusion it is automatically true that $i(n^*n) = i(n)^*i(n)$ thus, for any $n \in N$ we have
\beq \label{GCE homogeneity}
	i^{\dagger}_{\varphi} \circ i(n) = n \iff i \circ \sigma^{\varphi \circ i}_t(n) = \sigma^{\varphi}_t \circ i(n).
\eeq
So $N_{i,\varphi} \subset N$ is the set of elements for which $\sigma^{\varphi \circ i}_t = \sigma^{\varphi}_t \rvert_{N}$ for all $t$. If Eqn.~\eqref{GCE homogeneity} holds, it can moreover be shown that
\beq
	i^{\dagger}_{\varphi}(i(n) \; m \; i(n)^*) = n \; i^{\dagger}_{\varphi}(m) \; n^*, \; \forall m \in M, n \in N_{i,\varphi}. 
\eeq	
A corollary of this observation is that $i^{\dagger}_{\varphi}: M \rightarrow N$ is a conditional expectation if and only if $N_{i,\varphi} = N$. This reproduces a theorem by Takesaki \cite{takesaki_conditional_1972}. More generally, if there is a restriction of the map $i^{\dagger}_{\varphi}$ with image  $N_{i,\varphi}$, then this restriction is a conditional expectation.  Note that this establishes an immediate connection with \textit{approximate} error correction and generalized conditional expectations. Unless $i^{\dagger}_{\varphi}$ is a conditional expectation, one cannot recover the entire subalgebra $N$.

Let us concentrate for the moment on the case that $i_{\varphi}^{\dagger}$ is a conditional expectation and the modularity condition in Eqn.~\eqref{GCE homogeneity} holds for every $n \in N$. A corollary \cite{takesaki_conditional_1972} of the modularity condition is that the von Neumann algebra generated by $i(N)$ and its relative commutant $i(N)^c\equiv \{m \in M \; | \; [m,i(n)] = 0, \; \forall n \in N\}$ tensor factorizes: 
\beq
    i \circ \sigma^{\varphi \circ i}_t = \sigma^{\varphi}_t \circ i \iff i(N) \vee i(N)^c \simeq i(N) \otimes i(N)^c.
\eeq
If $i: N \hookrightarrow M$ is a co-normal inclusion, meaning $M \simeq i(N) \vee i(N)^c$, the existence of a conditional expectation therefore implies a tensor factorization of the algebra

\begin{equation}
    M \simeq i(N) \vee i(N)^{c} \simeq i(N) \otimes i(N)^{c}.
\end{equation}
From this perspective, we regard the generalized conditional expectation as a device for factorizing states, $\varphi = \varphi_0 \circ i_{\varphi}^{\dagger}$, \textit{without} requiring a tensor factorization of the algebra. As we will discuss in Section \ref{sec: DPI}, this observation is crucial to constructing a fully quantum notion of conditional entropy. 

The previous theorem can be regarded as a statement of \textit{operator} algebra error correction. Given a channel $\alpha: N \rightarrow M$ the subalgebra $N_{\alpha,\varphi} \subset N$ is the largest algebra that can be recovered (relative to the Petz map induced by the input state $\varphi$). In a similar vein, we can ask the recoverability question but about states. The analogue of recoverability becomes sufficiency, namely the ability to distinguish two pairs of input states after passing them through the channel.
\begin{definition}[Sufficiency]
	Let $\alpha: N \rightarrow M$ be a channel and $\mathcal{F} \subset P(M)$ a family of states. We say that $\alpha$ is sufficient with respect to $\mathcal{F}$ if there exists a channel $\beta: M \rightarrow N$ such that $\varphi \circ \alpha \circ \beta = \varphi$ for all $\varphi \in \mathcal{F}$. 
\end{definition}
A seminal theorem by Petz establishes the following result:\footnote{Here $u^{\varphi \mid \psi}_t \in M$ is the Connes' cocycle derivative, and $D(\varphi \parallel \psi) \equiv i \lim_{t \rightarrow 0} t^{-1} \varphi(u^{\varphi \mid \psi}_t - \mathbb{1}_M)$ is the relative entropy. We define these more carefully in Section \ref{sec: Spatial Theory}.}
\begin{theorem}[Sufficiency]
Let $\alpha: N \rightarrow M$ and $\mathcal{F} \equiv \{\varphi,\psi\} \subset P(M)$. The following are equivalent:
\begin{enumerate}
	\item $\alpha \circ \alpha^{\dagger}_{\psi}(u^{\varphi \mid \psi}_t) = u^{\varphi \mid \psi}_t, \; \forall t \in \mathbb{R}$, 
	\item $D(\varphi \circ \alpha \parallel \psi \circ \alpha) = D(\varphi \mid \psi)$, 
	\item $\alpha(u^{\varphi \circ \alpha \mid \psi \circ \alpha}_t) = u^{\varphi \mid \psi}_t, \; \forall t \in \mathbb{R}$,
	\item $\alpha^{\dagger}_{\varphi} = \alpha^{\dagger}_{\psi}$,
	\item $\varphi \circ \alpha \circ \alpha^{\dagger}_{\psi} = \varphi$, 
	\item $\psi \circ \alpha \circ \alpha^{\dagger}_{\varphi} = \psi$. 
\end{enumerate}
\end{theorem}

Among other things, this theorem establishes that $\alpha$ is sufficient with respect to $\{\varphi,\psi\} \subset P(M)$ if and only if $\varphi$ and $\psi$ saturate the data processing inequality under the channel $\alpha$, which is the second condition of the above theorem. This is what is meant by the statement that $\varphi$ and $\psi$ remain distinguishable under $\alpha$. The final three conditions of the previous theorem imply that $\alpha_{\mathcal{F}}^{\dagger} \equiv \alpha^{\dagger}_{\varphi} = \alpha^{\dagger}_{\psi}$ is a recovery map\footnote{In general, one can always use $\alpha^{\dagger}_{\varphi}$ to recover expectation values of $\varphi$ under $\alpha$ -- $\varphi \circ \alpha \circ \alpha^{\dagger}_{\varphi} = \varphi$. However, it is only true that $\alpha^{\dagger}_{\varphi}$ can be used to protect the expectation values of another state $\psi \neq \varphi \in P(M)$ if $\varphi$ and $\psi$ remain distinguishable under the channel.}
\beq
	\varphi \circ \alpha \circ \alpha^{\dagger}_{\mathcal{F}} = \varphi, \; \forall \varphi \in \mathcal{F}. 
\eeq

It can be shown that $\alpha$ is sufficient with respect to a family $\mathcal{F} \subset P(M)$ if and only if it is sufficient with respect to each pair $\{\varphi,\psi\} \subset \mathcal{F}$. Given a subalgebra $N \subset M$, we say that $N$ is sufficient with respect to $\mathcal{F}$ if $i: N \hookrightarrow M$ is sufficient as a channel. If a subalgebra is sufficient with respect to a family of states $\mathcal{F}$, one gains no improvement in the ability to distinguish these states by observing operators in $M$ outside of $N$. In other words, all of the information in $M$ which is relevant for distinguishing between elements of $\mathcal{F}$ is contained in $N \subset M$. 

From Petz's theorem we conclude that, given a family $\mathcal{F} \subset P(M)$ the smallest sufficient subalgebra for $\mathcal{F}$ is
\beq
	N_{\mathcal{F}} \equiv \{u^{\varphi \mid \psi}_t \; | \; t \in \mathbb{R}, \{\varphi,\psi\} \subset \mathcal{F}\}. 
\eeq
Then, $N \subset M$ is sufficient for $\mathcal{F}$ if and only if it contains $N_\mathcal{F}$. In a similar vein, given a channel $\alpha: N \rightarrow M$ we define the subalgebra
\beq
	M_{\alpha,\mathcal{F}} \equiv \{m \in M \; | \; \alpha \circ \alpha^{\dagger}_{\varphi}(m) = m, \; \forall \varphi \in \mathcal{F}\}. 
\eeq
We see that $\alpha$ is sufficient for $\mathcal{F}$ as a channel if and only if $M_{\alpha,\mathcal{F}} \subset M$ is sufficient for $\mathcal{F}$ as a subalgebra, e.g. if and only if $N_{\mathcal{F}} \subset M_{\alpha,\mathcal{F}}$. In this way, the study of quantum error correction for generic quantum channels can always be reduced to that of inclusions of operator algebras. 

\subsection{Spatial Theory} \label{sec: Spatial Theory}
Spatial theory is the cornerstone of Connes' approach \cite{CONNES1980153} to Tomita and Takesaki's modular theory \cite{Takesaki:1970aki}. At the level of weights on von Neumann algebras, spatial theory can be interpreted as a non-commutative generalization of Radon-Nikodym theory. Let $M$ be a von Neumann algebra and $\varphi \in P(M)$ a faithful, semifinite, normal weight. Denote by $L^2(M;\varphi)$ the GNS Hilbert space of $M$ with respect to $\varphi$, and by $\xi_{\varphi} \in L^2(M;\varphi)$ its vector representative. A weight $\psi \in M_*$ (not necessarily faithful, semifinite, and/or normal) is called differentiable with respect to $\varphi$ if there exists a vector $\xi_{\psi} \in L^2(M;\varphi)$ such that
\beq
	\psi(m) = g_{L^2(M;\varphi)}(\xi_{\psi}, \pi_{\varphi}(m) \xi_{\psi}). 
\eeq
In this case, there exists a positive self-adjoint operator $\rho_{\psi \mid \varphi}^{1/2}$ affiliated with $\pi_{\varphi}(M)'$ such that $\xi_{\psi} = \rho_{\psi \mid \varphi}^{1/2} \xi_{\varphi}$. Essentially, $\rho_{\psi \mid \varphi}^{\frac{1}{2}}$ functions as a non-commutative Jacobian between two weights.

Given any $\psi \in P(M)$, this will always be the case and in fact\footnote{Here $\overline{\pi}_{\varphi}(m) \equiv J_{\varphi} \pi_{\varphi}(m^*) J_{\varphi}$ is the canonical antirepresentation of $M$ on the GNS Hilbert space. Note that $\overline{\pi}_{\varphi}(M) \simeq \pi_{\varphi}(M)'$.} $\rho_{\psi \mid \varphi}^{1/2} = \overline{\pi}_{\varphi}(u^{\psi \mid \varphi}_{-i/2})$ with
\beq
	u^{\psi \mid \varphi}_t = \pi_{\varphi}^{-1}\bigg(\Delta_{\psi \mid \varphi}^{it} \Delta_{\varphi}^{-it}\bigg) = \pi_{\varphi}^{-1} \bigg(\Delta_{\psi}^{it} \Delta_{\varphi \mid \psi}^{-it}\bigg),
\eeq
the Connes' cocycle derivative. The operator $\rho_{\psi \mid \varphi}$ is called the density operator of $\psi$ relative to $\varphi$, or the spatial derivative of $\psi$ with respect to $\varphi$.\footnote{It may also be identified as the one sided relative modular operator of $\psi$ with respect to $\varphi$.} It can be used to define the relative entropy between the weights $\psi$ and $\varphi$:\footnote{We have initiated the following notation: Given a Hilbert space $H$ and a vector $\xi \in H$ we define the weight $\omega^H_{\xi} \in B(H)_*$ by
\beq
    \omega^H_{\xi}(\mathcal{O}) \equiv g_H(\xi, \mathcal{O} \xi). 
\eeq}
\beq
	D(\psi \parallel \varphi) \equiv \omega_{\xi_{\psi}}^{L^2(M;\varphi)}\bigg(\ln \rho_{\psi \mid \varphi}\bigg). 
\eeq
In the event that $M$ is semi-finite with tracial weight $\tau \in P(M)$ the operator $\rho_{\psi \mid \tau}$ is called the density operator of $\psi$ with respect to the trace $\tau$. Such a density operator can be used to define a notion of von Neumann entropy, up to a possibly infinite but state-independent constant associated with the choice of $\tau$ \cite{longo_note_2022}
\beq \label{Semifinite Entropy}
    S_{\tau}(\psi) \equiv -D(\psi \parallel \tau). 
\eeq
If $M$ is at least type II$_1$ it admits a tracial \emph{state} (which is normalizable), and the density operator can be uniquely defined with respect to this state.

Notice that a state $\varphi: M \rightarrow \mathbb{C}$ may be interpreted as a conditional expectation, with $\mathbb{C} \mathbb{1} \subset M$ regarded as a subalgebra. Working in the GNS representation $L^2(M;\varphi)$, the vector representative $\xi_{\varphi}$ gives rise to an isometry $W_{\varphi \mid L^2(M;\varphi)}: \mathbb{C} \rightarrow L^2(M;\varphi)$ with $W_{\varphi \mid L^2(M;\varphi)}(z) \equiv z \xi_{\varphi}$. The adjoint map is implemented by the dual vector as $W_{\varphi \mid L^2(M;\varphi)}^{\dagger}(\eta_{\varphi}(m)) = g_{\varphi}(\xi_{\varphi}, \eta_{\varphi}(m)) = \varphi(m)$. Thus, we see that
\beq
	W_{\varphi \mid L^2(M;\varphi)}^{\dagger} \pi_{\varphi}(m) W_{\varphi \mid L^2(M;\varphi)}(z) = W_{\varphi \mid L^2(M;\varphi)}^{\dagger}(z \eta_{\varphi}(m)) = \varphi(m) z,
\eeq
or $W_{\varphi \mid L^2(M;\varphi)}^{\dagger} \pi_{\varphi}(m) W_{\varphi \mid L^2(M;\varphi)} = \varphi(m)$. In this sense, a state is automatically spatial in its GNS representation when viewed as a completely positive map. By construction, the relative density operator $\rho_{\psi \mid \varphi}^{1/2}$ satisfies the equation
\beq
	\psi(m) = \bigg(\rho_{\psi \mid \varphi}^{1/2} W_{\varphi \mid L^2(M;\varphi)}\bigg)^{\dagger} \pi_{\varphi}(m) \bigg(\rho_{\psi \mid \varphi}^{1/2} W_{\varphi \mid L^2(M;\varphi)}\bigg) \equiv W_{\psi \mid L^2(M;\varphi)}^{\dagger} \pi_{\varphi}(m) W_{\psi \mid L^2(M;\varphi)}.
\eeq
That is, the existence of $\rho_{\psi \mid \varphi}^{1/2}$ establishes the spatial nature of $\psi$ with respect to the GNS Hilbert space of $\varphi$, with spatial implementation 
\beq
	W_{\psi \mid L^2(M;\varphi)} = \rho_{\psi \mid \varphi}^{1/2} W_{\varphi \mid L^2(M;\varphi)}. 
\eeq

In \cite{BELAVKIN198649}, a spatial theory is constructed for completely positive maps which reduces to Connes' spatial theory in the event that the maps in question are weights. Let $\alpha,\beta: A \rightarrow B(H)$ be two completely positive maps and suppose that $\beta$ is spatial with respect to the representation $\pi_{K}: A \rightarrow B(K)$. That is, there exists $W_{\beta \mid K}: H \rightarrow K$ such that $\beta(a) = W_{\beta \mid K}^{\dagger} \pi_K(a) W_{\beta \mid K}$. Then, $\alpha$ is called differentiable\footnote{In \cite{BELAVKIN198649} it is shown that this property is equivalent to the statement that $\alpha$ is strongly, completely, absolutely continuous with respect to $\beta$.} with respect to $\beta$ if and only if it is spatial with respect to $\pi_K$ and there exists a positive, self-adjoint operator affiliated with $\pi_K(A)'$ such that
\beq
	\alpha(a) = \bigg(\rho_{\alpha \mid \beta}^{1/2} W_{\beta \mid K}\bigg)^{\dagger} \pi_K(a) \bigg(\rho_{\alpha \mid \beta}^{1/2} W_{\beta \mid K}\bigg) \equiv W_{\alpha \mid K}^{\dagger} \pi_K(a) W_{\alpha \mid K}. 
\eeq
The map $\rho_{\alpha \mid \beta}: K \rightarrow K$ is called the density operator of $\alpha$ relative to $\beta$, or the spatial derivative of $\alpha$ by $\beta$. We see that in the case where one of the maps $\beta$ is spatial with $W_{\beta|K}$ as its implementer, the differentiability of the other $\alpha$ with respect to it allows us to obtain the implementer $W_{\alpha|K}$ of $\alpha$ by using the `Jacobian' $\rho_{\alpha|\beta}$.

\section{Factorization of quantum entropy} \label{sec: DPI}

Our aim is to discover quantum analogs for the equations Eqs.~\eqref{Classical Factorization}, \eqref{Classical Generalized Entropy}, and \eqref{Classical Data Processing 2}. As we have stressed in Section \ref{sec: classicalrevi}, the latter observations follow immediately from the first which is Bayes' law. Thus, our goal is to derive a quantum Bayes' law and then to derive the consequences of this law on the relative and von Neumann entropy of states. To this end, we use the machinery presented in Section \ref{sec: CP Maps} to construct a quantum Bayes' law which is valid for \textit{any} von Neumann algebraic inclusion independently of assumptions pertaining to semifiniteness or tensor factorizability.
\subsection{Quantum Bayes' law}

To begin, let us consider how the relative density operator factorizes over the composition of completely positive maps. Take $\alpha_1,\beta_1: A_1 \rightarrow B(H_2)$ and $\alpha_2,\beta_2: B(H_2) \rightarrow B(H_3)$ to be completely positive maps so that their compositions, $\alpha_3 \equiv \alpha_2 \circ \alpha_1, \beta_3 \equiv \beta_2 \circ \beta_1: A_1 \rightarrow B(H_3)$ are also completely positive. Suppose that $\beta_i$ are all spatial with respect to a common Hilbert space $K$ with $\pi^1_K: A_1 \rightarrow B(K)$, and $\pi^2_K: B(H_2) \rightarrow B(K)$ such that
\begin{flalign}
	&\beta_1(a_1) = W_{\beta_1 \mid K}^{\dagger} \pi^1_K(a_1) W_{\beta_1 \mid K}, \nonumber \\
	&\beta_2(\mathcal{O}_2) = W_{\beta_2 \mid K}^{\dagger} \pi^2_K(\mathcal{O}_2) W_{\beta_2 \mid K}, \nonumber \\
	&\beta_3(a_1) = W_{\beta_3 \mid K}^{\dagger} \pi^1_K(a_1) W_{\beta_3 \mid K}.
\end{flalign}
Here, $W_{\beta_1 \mid K}: H_2 \rightarrow K$, $W_{\beta_2 \mid K}: H_3 \rightarrow K$, and $W_{\beta_3 \mid K}: H_3 \rightarrow K$ are the spatial implementers of the maps $\beta_{i}$ on $K$. Since the map $\pi^2_K$ is a representation mapping one set of bounded operators to another, it can be implemented by an isometry $U: H_2 \rightarrow K$ such that $\pi_2^K(\mathcal{O}_2) = U \mathcal{O}_2 U^{\dagger}$. 

Now, suppose that each $\alpha_i$ is differentiable with respect to $\beta_i$. Then, we get relative density operators $\rho_{\alpha_i \mid \beta_i}: K \rightarrow K$ such that
\begin{flalign} \label{Spatial reps of alphas}
	&\alpha_1(a_1) = \bigg(\rho_{\alpha_1 \mid \beta_1}^{1/2} W_{\beta_1 \mid K}\bigg)^{\dagger} \pi^1_K(a_1) \bigg(\rho_{\alpha_1 \mid \beta_1}^{1/2} W_{\beta_1 \mid K}\bigg), \nonumber \\
	&\alpha_2(\mathcal{O}_2) = \bigg(\rho_{\alpha_2 \mid \beta_2}^{1/2} W_{\beta_2 \mid K}\bigg)^{\dagger} \pi^2_K(\mathcal{O}_2) \bigg(\rho_{\alpha_2 \mid \beta_2}^{1/2} W_{\beta_2 \mid K}\bigg), \nonumber \\
	&\alpha_3(a_1) = \bigg(\rho_{\alpha_3 \mid \beta_3}^{1/2} W_{\beta_3 \mid K}\bigg)^{\dagger} \pi^1_K(a_1) \bigg(\rho_{\alpha_3 \mid \beta_3}^{1/2} W_{\beta_3 \mid K}\bigg).
\end{flalign}
We also know that the third channel can be obtained by taking the composition of the first two. In particular,
\beq \label{Composition part 2}
	\alpha_3(a_1) = \bigg(\rho_{\alpha_2 \mid \beta_2}^{1/2} W_{\beta_2 \mid K}\bigg)^{\dagger} U \bigg(\rho_{\alpha_1 \mid \beta_1}^{1/2} W_{\beta_1 \mid K}\bigg)^{\dagger} \pi_K^1(a_1) \bigg(\rho_{\alpha_1 \mid \beta_1}^{1/2} W_{\beta_1 \mid K}\bigg) U^{\dagger} \bigg(\rho_{\alpha_2 \mid \beta_2}^{1/2} W_{\beta_2 \mid K}\bigg).
\eeq
Equating Eqn.~\eqref{Composition part 2} with the third line of Eqn.~\eqref{Spatial reps of alphas} we find
\beq \label{General conditional relation}
	\rho_{\alpha_3 \mid \beta_3}^{1/2} W_{\beta_3 \mid K} = \rho_{\alpha_1 \mid \beta_1}^{1/2} W_{\beta_1 \mid K} U^{\dagger} \rho_{\alpha_2 \mid \beta_2}^{1/2} W_{\beta_2 \mid K}.,
\eeq
which can be interpreted as a chain rule for the spatial derivative. 

In the most general case, Eqn.\eqref{General conditional relation} is as far as we can go. However, in a case of particular interest we can massage Eqn.~\eqref{General conditional relation} further. Let $\alpha_3 \equiv \psi = \psi_0 \circ i_{\psi}^{\dagger}$ be a state on $M$ with $\psi_0 \equiv \psi \circ i$ a state on $N$ for $i: N \hookrightarrow M$ an inclusion. The map $i_{\psi}^{\dagger}: M \rightarrow N$ is the generalized conditional expectation induced by $\psi$. Similarly, let $\beta_3 \equiv \varphi = \varphi_0 \circ i_{\varphi}^{\dagger}$ be a state on $M$ with $i$ the same inclusion, and $i_{\varphi}^{\dagger}$ the generalized conditional expectation induced by $\varphi$. Each of the maps $\varphi: M \rightarrow \mathbb{C}$, $\varphi_0: N \rightarrow \mathbb{C}$ and $i_{\varphi}^{\dagger}: M \rightarrow N$ are spatial on the GNS Hilbert space $L^2(M;\varphi)$. In particular, 
\begin{flalign}
	&\varphi(m) = W_{\varphi \mid L^2(M;\varphi)}^{\dagger} \pi_{\varphi}(m) W_{\varphi \mid L^2(M;\varphi)}, \nonumber \\
	&\varphi_0(n) = W_{\varphi \mid L^2(M;\varphi)}^{\dagger} \pi_{\varphi} \circ i(n) W_{\varphi \mid L^2(M;\varphi)},
\end{flalign}
since $\varphi$ and $\varphi_0$ are implemented with respect to the same vector in $L^2(M;\varphi)$:
\beq
	\varphi(m) = g_{\varphi}(\xi_{\varphi}, \pi_{\varphi}(m) \xi_{\varphi}), \qquad \varphi_0(n) = \varphi \circ i(n) = g_{\varphi}(\xi_{\varphi}, \pi_{\varphi} \circ i(n) \xi_{\varphi}). 
\eeq
Assuming that $\psi$ is differentiable with respect to $\varphi$ and $\psi_0$ is differentiable with respect to $\varphi_0$, in this instance Eqn.~\eqref{General conditional relation} can be simplified to\footnote{From the discussion above, we can also write explicitly that
\beq
	W_{i_{\varphi}^{\dagger} \mid L^2(M;\varphi)} = J_{\varphi} V_i J_{\varphi_0},
\eeq
where $V_i$ implements the inclusion on the GNS Hilbert space.}
\beq
	\rho_{\psi \mid \varphi}^{1/2} = \rho_{i_{\psi}^{\dagger} \mid i_{\varphi}^{\dagger}}^{1/2} W_{i_{\varphi}^{\dagger} \mid L^2(M;\varphi)} U^{\dagger} \rho_{\psi_0 \mid \varphi_0}^{1/2}.
\eeq

Squaring this operator, we obtain an explicit formula relating the relative density operator between states on $M$ with the relative density operator for their restrictions to $N$:
\beq \label{Non-Commutative Conditional Factorization}
	\rho_{\psi \mid \varphi} = \bigg(U W_{i_{\varphi}^{\dagger} \mid L^2(M;\varphi)}^{\dagger} \rho_{i_{\psi}^{\dagger} \mid i_{\varphi}^{\dagger}}^{1/2}\bigg)^{\dagger} \rho_{\psi_0 \mid \varphi_0} \bigg(U W_{i_{\varphi}^{\dagger} \mid L^2(M;\varphi)}^{\dagger} \rho_{i_{\psi}^{\dagger} \mid i_{\varphi}^{\dagger}}^{1/2}\bigg) \equiv \bigg(\rho_{\psi \mid i,\varphi}^{1/2}\bigg)^{\dagger} \rho_{\psi_0 \mid \varphi_0} \rho_{\psi \mid i,\varphi}^{1/2}. 
\eeq
The above equation is the non-commutative version of Eqn.~\eqref{Classical Factorization}, a quantum Bayes' law. We refer to $\rho_{\psi_0 \mid \varphi_0}^{1/2}$ as the relative marginal density operator and $\rho_{\psi \mid i,\varphi}^{1/2}$ as the relative conditional density operator. In fact, Eqn.~\eqref{Non-Commutative Conditional Factorization} is valid for any quantum channel $\alpha: N \rightarrow M$ with:
\beq \label{Density factorization for channels}
    \rho_{\varphi \mid \psi} = \bigg(U W_{\alpha_{\varphi}^{\dagger} \mid L^2(M;\varphi)}^{\dagger} \rho_{\alpha_{\psi}^{\dagger} \mid \alpha_{\varphi}^{\dagger}}^{1/2}\bigg)^{\dagger} \rho_{\psi \circ \alpha \mid \varphi \circ \alpha} \bigg(U W_{\alpha_{\varphi}^{\dagger} \mid L^2(M;\varphi)}^{\dagger} \rho_{\alpha_{\psi}^{\dagger} \mid \alpha_{\varphi}^{\dagger}}^{1/2}\bigg) \equiv \bigg(\rho_{\alpha_{\psi}^{\dagger} \mid \alpha,\varphi}^{1/2}\bigg)^{\dagger} \rho_{\psi \circ \alpha \mid \varphi \circ \alpha} \; \rho_{\alpha_{\psi}^{\dagger} \mid \alpha,\varphi}^{1/2},
\eeq
by similar manipulations.
\subsection{Data processing inequality}
An immediate corollary of Eqn.~\eqref{Density factorization for channels} is that it allows us to derive an expression relating the relative entropy of states with the relative entropy of the same states under the application of a quantum channel:
\beq \label{Quantification of DPI}
    D(\psi \parallel \varphi) = D(\psi \circ \alpha \parallel \varphi \circ \alpha) + \omega_{\xi_{\psi}}^{L^2(M;\varphi)}\bigg(\ln \rho_{\alpha_{\psi}^{\dagger} \mid \alpha,\varphi}\bigg) + \sum_{n = 1}^{\infty} \frac{c_n}{n!} \omega_{\xi_{\psi}}^{L^2(M;\varphi)} \bigg(\text{ad}_{\ln \rho_{\alpha_{\psi}^{\dagger} \mid \alpha,\varphi}}^n(\ln \rho_{\psi \circ \alpha \mid \varphi \circ \alpha}) \bigg) 
\eeq
where $\text{ad}_{a}^n(b)$ is the $n$-fold commutator of $a$ and $b$, and $c_n$ are constants. Eqn.~\eqref{Quantification of DPI} is the quantum analog of \eqref{Classical Data Processing 1}. The first term arises from the relative marginal density operator, the second from the relative conditional marginal density operator, and the remaining terms from their lack of commutativity. These additional terms can be derived precisely using the Baker-Campell-Hausdorff formula, and do not arise in the classical setting because the marginal and conditional density `operators' will always commute in that case. 

As was true in the classical setting, Eqn.~\eqref{Quantification of DPI} may be regarded as a quantification of the data processing inequality:
\beq
    D(\psi \parallel \varphi) - D(\psi \circ \alpha \parallel \varphi \circ \alpha) \geq 0.
\eeq
Per Petz's theorem, the data processing inequality is saturated if and only if $\alpha$ is sufficient for $\{\psi,\varphi\}$. We may therefore regard the series of terms associated with the quantum conditional entropy as encoding obstructions to the sufficiency of $\alpha$ with respect to a given pair of states.

In general, there is no analogue of Eqn.~\eqref{Classical Generalized Entropy} as the von Neumann entropy is ill-defined unless the von Neumann algebra in question is semi-finite. In the event that $M$ is a semi-finite algebra admitting a tracial weight $\tau \in P(M)$ we can use \eqref{Non-Commutative Conditional Factorization} to compute the entropy of a generic state $\varphi \in S(M)$ by evoking \eqref{Semifinite Entropy}. In particular, we find that
\beq \label{General Generalized Entropy Formula}
    S_{\tau}(\varphi) = -D(\varphi \parallel \tau \circ i) + \omega^{L^2(M;\tau)}_{\xi_{\varphi}}\bigg(-\ln \rho_{\varphi \mid i,\tau} + ... \bigg) \equiv -D(\varphi \parallel \tau \circ i) + \varphi_0( A_{i_{\varphi}^{\dagger}}). 
\eeq
Here, we have defined the operator $A_{i_{\varphi}^{\dagger}}$ associated with the generalized conditional expectation $i_{\varphi}^{\dagger}$ and labeled it $A$ in anticipation of its role as an area in the gravitational context. This operator is generically non-central, except in the case that $i_{\varphi}^{\dagger}$ is a conditional expectation -- as we shall discuss shortly. 

We should also emphasize that we have not assumed $N \subset M$ is a semi-finite algebra. When this is the case, we can choose $\tau$ so that $\tau_0 \equiv \tau \circ i \in P(N)$ is also a faithful, semi-finite, normal trace. Then, the first term on the right hand side of \eqref{General Generalized Entropy Formula} can be interpreted as the entropy of the marginalized state $\varphi_0$ in $N$. On the other hand, if $N$ is \emph{not} semi-finite, $\tau \circ i$ is merely a completely positive map on $N$, rather than a weight. In this case, the first term on the right hand side of \eqref{General Generalized Entropy Formula} is not a true entropy, and contains a divergent contribution. Nevertheless, as we expect the full entropy $S_{\tau}(\varphi)$ to be finite (up to a state independent divergent constant) we conclude that there must be a cancellation between the divergent `entropy' of the state on the algebra $N$ and the expectation value of the area operator. Such a cancellation can be interpreted as a manifestation of the Susskind-Uglum conjecture~\cite{Susskind_1994}. This state of affairs is of particular interest to developing a physical interpretation of the entropy of states as computed in the modular crossed product of a QFT subregion algebra, as we discuss below.

\section{The Area as Quantum Conditional Entropy} \label{sec: applications}

In this section, we point our non-commutative Bayes' theorem towards a handful of different physical and mathematical applications. The main thread of analysis centers around the generalized entropy. In particular, we demonstrate how the area term in the generalized entropy formula can be interpreted as a quantum conditional entropy in both the exact and approximate quantum error correcting regimes. This connects the dots between the original algebraic approach to the generalized entropy in terms of quantum error correction and more recent perspectives which stem from the modular crossed product. 

We formulate our presentation as a series of examples of increasing `quantum-ness'. We begin by recasting our classical derivation in operator algebraic language, then discuss the case of exact quantum error correction in which algebraic factorization suppresses quantum entanglement between the area and the rest of the algebra, and finally treat the case of the modular crossed product. This series of computations serves to illustrate how these apparently disparate ideas fit neatly together from the point of view of quantum conditional probability.

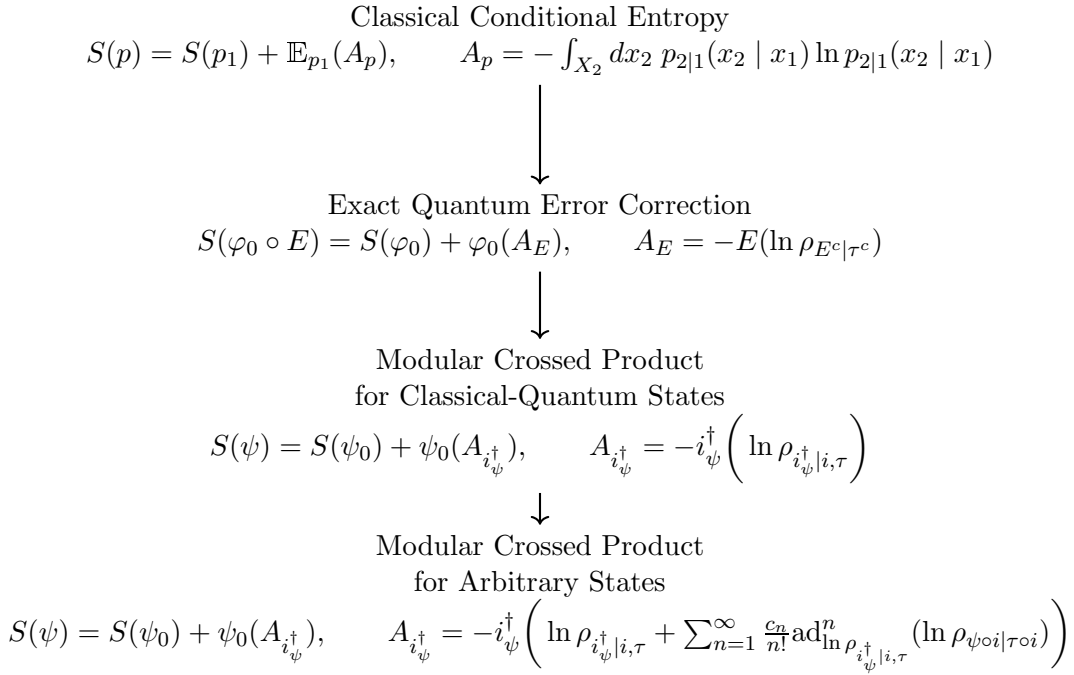
\begin{figure}[h!]
    \centering
    \begin{tikzpicture}[node distance=2.5cm, align=center]
        \node (A) {Classical Conditional Entropy \\
        $S(p) = S(p_1) + \mathbb{E}_{p_1}(A_p), \qquad A_p = -\int_{X_2} dx_2 \; p_{2 \mid 1}(x_2 \mid x_1) \ln p_{2 \mid 1}(x_2 \mid x_1)$};
        \node (B) [below of=A] {Exact Quantum Error Correction \\
        $S(\varphi_0 \circ E) = S(\varphi_0) + \varphi_0(A_E), \qquad A_E = -E(\ln \rho_{E^c \mid \tau^c})$};
        \node (C) [below of=B] {Modular Crossed Product \\ for Classical-Quantum States \\
        $S(\psi) = S(\psi_0) + \psi_0(A_{i_{\psi}^{\dagger}}), \qquad A_{i_{\psi}^{\dagger}} = -i_{\psi}^{\dagger}\bigg(\ln \rho_{i_{\psi}^{\dagger} \mid i,\tau}\bigg)$};
        \node (D) [below of=C] {Modular Crossed Product \\ for Arbitrary States \\
        $S(\psi) = S(\psi_0) + \psi_0(A_{i_{\psi}^{\dagger}}), \qquad A_{i_{\psi}^{\dagger}} = -i_{\psi}^{\dagger}\bigg(\ln \rho_{i_{\psi}^{\dagger} \mid i,\tau} + \sum_{n = 1}^{\infty} \frac{c_n}{n!} \text{ad}_{\ln \rho_{i_{\psi}^{\dagger} \mid i,\tau}}^n(\ln \rho_{\psi \circ i \mid \tau \circ i}) \bigg) $};

        \draw[->, thick] (A) -- (B);
        \draw[->, thick] (B) -- (C);
        \draw[->, thick] (C) -- (D);
    \end{tikzpicture}
    \caption{Sequence of computations.}
\end{figure}

\subsection{Classical Case} \label{sec: Classical}

As a first exercise, it is quite helpful to recontextualize the classical example of conditional probability theory in our spatial language. Classical probability theory can be formulated in terms of a commutative von Neumann algebra. Again, let $X = X_1 \times X_2$ be a measure space. The commutative algebra of operators associated with $X$ is its function space $L^{\infty}(X)$ along with the pointwise product of functions.\footnote{For simplicity, we will work with real valued functions so that the involution is trivial. In the complex case it would be conjugation.} The weight space of this algebra is its predual, $L^{\infty}(X)_* = L^{1}(X)$. A weight on $L^{\infty}(X)$ is therefore a map $p: X \rightarrow \mathbb{R}$ such that
\beq
	\tr(p) \equiv \int dx_1 \; dx_2 \; p(x_1,x_2) < \infty. 
\eeq	
Any such map can be normalized to obtain a state, which is an element of $L^1(X)$ for which the above integral is one. Here, we have enlisted the notation $\tr$ for the integral over $X$ with respect to its Lebesgue measure to indicate that this should be interpreted as a trace.\footnote{Note that \emph{any} element of $L^1(X)$ is in fact a tracial weight, since the algebra in question is commutative. The notation here is simply meant to identify the integral against the standard Lebesgue measure as a reference map.} Indeed, $\tr: L^{\infty}(X) \rightarrow \mathbb{R}$ may itself be thought of as a completely positive linear functional on $L^{\infty}(X)$, although it may not belong explicitly to the predual if the space is not compact. To avoid unnecessary subtlety, we will assume that $X$ is compact and that the trace is suitably normalized.

Each element $p \in L^1(X)$ automatically defines a map $\mathbb{E}_p: L^{\infty}(X) \rightarrow \mathbb{R}$ by
\beq \label{Expectation value}
	\mathbb{E}_p(f) = \tr(p f) = \int_{X} dx_1 \; dx_2 \; p(x_1,x_2) f(x_1,x_2),
\eeq 
which we recognize as the expectation value of $f: X \rightarrow \mathbb{R}$, regarded as a random variable, with respect to the distribution defined by $p$. In this respect, Eqn.~\eqref{Expectation value} identifies $p$ as the `density operator' associated with the weight $\mathbb{E}_p$. More precisely, in the language we have developed above we see that $p$ is the (square of the) spatial derivative of the expectation functional with respect to the trace: $p = \rho_{\mathbb{E}_p \mid tr}$. It's worth noting that the spatial derivative completely reduces to the Radon-Nikodym derivative in the commutative context. Thus we may also regard $\rho_{\mathbb{E}_p \mid tr} = \frac{d \mu_p}{d \mu_X}$, with $d\mu_X(x_1,x_2) = dx_1 dx_2$ the standard Lebesgue measure on $X$ and $d\mu_p(x) = p(x) d\mu_X(x)$ the Radon measure associated with $p$.

Since we took $X = X_1 \times X_2$ we also have that $L^{\infty}(X) = L^{\infty}(X_1) \otimes L^{\infty}(X_2)$. We may therefore define the algebraic inclusion $i_1: L^{\infty}(X_1) \hookrightarrow L^{\infty}(X)$ such that $i_1(f)\rvert_{(x_1,x_2)} = f(x_1)$, and likewise for $i_2$. For each $f \in L^{\infty}(X_1)$, we have:
\begin{flalign} \label{Marginal Distribution}
	\mathbb{E}_p \circ i_1(f) = \int_{X_1} dx_1 \bigg(\int_{X_2} dx_2 \; p(x_1,x_2) \bigg) f(x_1) = \tr_1\bigg(\tr_{2 \mid 1}(p) f\bigg) \equiv \mathbb{E}_{p_1}(f).
\end{flalign}
Here $\tr_1(f) \equiv tr \circ i_1(f) = \int_{X_1} dx_1 \; f(x_1)$ is the `marginal trace' on $L^{\infty}(X_1)$ and $tr_{2 \mid 1}(p) \equiv \int_{X_2} dx_2 \; p(x_1,x_2)$ is the `partial trace' mapping elements of $L^{\infty}(X)$ to $L^{\infty}(X_1)$. Clearly, $p_1 \equiv \tr_{2 \mid 1}(p)$ is precisely the marginal distribution of $p$ when restricted to $X_1$. The map $\mathbb{E}_{p_1}(f)$ is the expectation value of $f$ regarded as a random variable on $X_1$ with respect to the marginal density. From Eqn.~\eqref{Marginal Distribution} we also see that $p_1 = \rho_{\mathbb{E}_{p} \circ i_1 \mid tr \circ i_1}$.

Let $p_{2 \mid 1}(x_2 \mid x_1) \equiv p(x_1,x_2) p_1(x_1)^{-1}$ denote the conditional distribution. As a probability distribution on $X_2$ for each fixed $x_1 \in X_1$, the conditional distribution gives rise to a map $\mathbb{E}_{p \mid X_1}: L^{\infty}(X) \rightarrow L^{\infty}(X_1)$ as
\beq \label{Conditional Expectation}
	\mathbb{E}_{p \mid X_1}(f)\rvert_{x_1} \equiv \tr_{2 \mid 1}(p_{2 \mid 1} f)\rvert_{x_1}. 
\eeq
Eqn.~\eqref{Conditional Expectation} is the conditional expectation of probability theory. We may read this equation as designating the conditional distribution as the density operator associated with the conditional expectation induced by $p$ with respect to the partial trace. More succinctly, $p_{2 \mid 1} = \rho_{\mathbb{E}_{p \mid X_1} \mid tr_{2 \mid 1}}$. It is also worth noticing that $\mathbb{E}_{p \mid X_1}$ can be regarded as `inverting' the inclusion $i_1$:
\beq
	\mathbb{E}_{p \mid X_1} \circ i_1(f)\rvert_{x_1} = \int_{X_2} dx_2 \; p_{2 \mid 1}(x_2 \mid x_1) f(x_1) = f(x_1). 
\eeq
In this respect, $E_{p \mid X_1} \equiv i_{1,\mathbb{E}_p}^{\dagger}$ can be interpreted as the Petz dual of $i_1$ when viewed as a classical channel. In the same vein, we can write $tr_{2 \mid 1} = i_{1,tr}^{\dagger}$. Thus, the conditional density is given by the spatial derivative $p_{2 \mid 1} = \rho_{i_{1,\mathbb{E}_p}^{\dagger} \mid i_{1,tr}^{\dagger}}$.

Putting together Eqs.~\eqref{Expectation value} and \eqref{Conditional Expectation} we notice that any expectation value on $L^{\infty}(X)$ can be factorized as
\beq
	\mathbb{E}_p = \mathbb{E}_{p_1} \circ \mathbb{E}_{p \mid X_1} = \mathbb{E}_{p_2} \circ \mathbb{E}_{p \mid X_2}. 
\eeq
This is Bayes' law. Using the insights we have addressed about the associated (relative) density operators this further implies:
\beq
    \rho_{\mathbb{E}_p \mid tr} = \rho_{\mathbb{E}_{p} \circ i_1 \mid tr \circ i_1} \rho_{i_{1,\mathbb{E}_p}^{\dagger} \mid i_{1,tr}^{\dagger}},
\eeq
which is the commutative form of \eqref{Non-Commutative Conditional Factorization}.

\subsection{Exact quantum error correction} \label{sec: Exact}
Another useful subcase of Eqn.~\eqref{Non-Commutative Conditional Factorization} is exact error correction. In many ways, exact error correction has the complexion of classical-quantum conditional probability. This is because the existence of strict conditional expectations, as are implied by exact error correction, leads to a large degree of tensor factorization which diminishes quantum effects like entanglement.

Let $i: N \hookrightarrow M$ be a co-normal inclusion\footnote{Co-normality implies that $M \simeq i(N) \vee i(N)^c$. Here, $i(N)^c \equiv \{m \in M \; | \; [m,i(n)] = 0, \; \forall n \in N\}$ is the relative commutant of $i(N) \subset M$.} of semi-finite factors, and suppose that there exists a conditional expectation $E: M \rightarrow N$. In fact, this discussion can be extended to the context of arbitrary semi-finite von Neumann algebras with the only necessary assumption being the existence of a conditional expectation. We can also relax the assumption that $i$ is co-normal. However, the present set-up is sufficient to illustrate the simplifications to Eqn.~\eqref{Non-Commutative Conditional Factorization} which arise from the existence of a conditional expectation, and our above analysis covers the more general cases (in additional to cases even more general than those admitting conditional expectations).\footnote{A more general computation in which $N$ is allowed to have a non-trivial center is covered in \cite{faulkner2020holographic}.}

The existence of a conditional expectation implies that operators $n \in N$ can be exactly recovered when embedded into $M$. In particular, $E \circ i(n) = n$ for all $n \in N$. At the level of states, we define the code-subspace relative to the conditional expectation as the set
\beq
	S_E(M) \equiv \{\varphi_0 \circ E \; | \; \varphi_0 \in S(N)\}.
\eeq
This set is referred to as a code-subspace since the projection property of the conditional expectation, $(i \circ E) \circ (i \circ E) = i \circ E$, ensures that $\varphi \circ i \circ E = \varphi$ for all $\varphi \in S_E(M)$. The conditional expectation is a quantum channel, and thus $S_E(M)$ is the set of states on $M$ which are preserved by this channel. As $E \circ i(n) = n$ it is also clear to see that $\varphi_0 \circ E \circ i(n) = \varphi_0(n)$ for all $\varphi_0 \in S(N)$. 

As we have reviewed in Section \ref{sec: Error Correction}, a conditional expectation is a generalized conditional expectation satisfying the additional modularity property
\beq \label{CE Modularity}
	\sigma^{\varphi}_{t}\rvert_N = \sigma^{\varphi \circ i}_t, \; \forall \varphi \in S_E(M).
\eeq
Per Takesaki's theorem, Eqn.~\eqref{CE Modularity} holds if and only if $i(N) \vee i(N)^c \simeq i(N) \otimes i(N)^c$. By the assumption of co-normality, the existence of a conditional expectation therefore implies a tensor factorization of the algebra, $M \simeq i(N) \otimes i(N)^c$.

Given this factorization, a generic element in $M$ is given by $m = i(n) n^c$ where $n \in N$ and $n^c \in i(N)^c$. Applying the conditional expectation to such an element, we find
\beq
	E(i(n) n^c) = n E\rvert_{i(N)^c}(n^c).
\eeq
The homogeneity of the conditional expectation requires that the image of $E\rvert_{i(N)^c}$ be the center of $N$, $Z(N)$. In particular,
\beq
	n E\vert_{i(N)^c}(n^c) = E(i(n) n^c) = E(n^c i(n)) = E\rvert_{i(N)^c}(n^c) n, \; \forall n \in N.  
\eeq
So, in our simplified case with $N$ a factor, we find that $E^c \equiv E\rvert_{i(N)^c}: i(N)^c \rightarrow \mathbb{C}$ is in fact a state on $i(N)^c$. This observation allows us to write $\varphi \in S_E(M)$ as a product state:
\beq
	\varphi_0 \circ E(i(n) n^c) = \varphi_0(n) E^c(n^c). 
\eeq

As we assumed our algebras are semi-finite, we work with a tracial state, $\tau \in S(M)$, whose restrictions $\tau_0 \equiv \tau \circ i \in S(N)$ and $\tau^c \equiv \tau\rvert_{i(N)^c}$ are also tracial. Then, the density operator associated with a weight $\varphi \in S_E(M)$ is given by
\beq \label{Code density operator}
	\rho_{\varphi_0 \circ E \mid \tau} = i(\rho_{\varphi_0 \mid \tau_0}) \rho_{E^c \mid \tau^c}. 
\eeq
Here, $\rho_{E^c \mid \tau^c} \in i(N)^c$ and thus we can rewrite Eqn.~\eqref{Code density operator} as
\beq \label{Code density operator 2}
	\rho_{\varphi_0 \circ E \mid \tau} = \rho_{E^c \mid \tau^c}^{1/2} i(\rho_{\varphi_0 \mid \tau_0}) \rho_{E^c \mid \tau^c}^{1/2}.
\eeq
Naturally, the density operator $\rho_{E^c \mid \tau^c}$ can also be interpreted as the spatial derivative of the conditional expectation $E$ by the partial trace map $\tau_{N \mid N^c}(i(n) \otimes n^c) = n \tau^c(n^c)$. Thus, we observe that Eqn.~\eqref{Code density operator 2} is another instantiation of Eqn.~\eqref{Non-Commutative Conditional Factorization}. The entropy of the state $\varphi_0 \circ E \in S_E(M)$ (with respect to the trace $\tau$) can be computed as
\beq
	S(\varphi_0 \circ E) = S(\varphi_0) + \varphi_0(A_E),
\eeq
where here $A_E = -E(\ln \rho_{E^c \mid \tau^c}) \in Z(N) \simeq \mathbb{C}$ is the `area operator' associated with the conditional expectation $E$.

In \cite{harlow_ryutakayanagi_2017,faulkner2020holographic} it was argued that one should interpret holography in terms of quantum error correcting codes. From this point of view we should regard $M$ as the operator algebra assigned to a subregion in a holographic boundary CFT and $N$ a dual subregion algebra in the bulk. The inclusion $i: N \hookrightarrow M$ embeds geometrical operators into the boundary which are in turn recovered by the conditional expectation $E: M \rightarrow N$. Unfortunately, it is also widely recognized\footnote{In fact, this was also acknowledged in the cited papers.} that exact quantum error correction is not likely to hold in full holography. Recently, an alternative approach to deriving the generalized entropy has emerged around a notion from von Neumann algebra theory called the modular crossed product~\cite{Witten:2021unn, Chandrasekaran:2022cip,Chandrasekaran:2022eqq, Jensen:2023yxy, AliAhmad:2023etg, Klinger:2023tgi, Kudler-Flam:2023hkl, ahmad2024semifinitevonneumannalgebras}. As we will demonstrate, the entropy factorization results we have derived apply equally well to the modular crossed product as they did to the exact error correcting case. In this regard, our approach provides a clear line between these two ideas and positions the modular crossed product approach to the generalized entropy as an approximate error correcting version of \cite{harlow_ryutakayanagi_2017,faulkner2020holographic}.


\subsection{The Modular Crossed Product} \label{sec: Modular}

Let $N$ be a type III$_1$ factor representing the QFT algebra of a finite subregion $\Sigma$.\footnote{To be clear, a number of physical assumptions must be met to justify the physicality of this construction. We refer the reader to \cite{Jensen:2023yxy} for an excellent discussion of these.} It has been argued that gravitational effects require the subregion algebra to be promoted to a crossed product $M \equiv N \times_{\sigma^{\varphi_0}} \mathbb{R}$ in order to maintain diffeomorphism covariance. Here $\varphi_0 \in S(N)$ is a reference state in the field theory, typically identified with the vacuum, and $\sigma^{\varphi_0}$ is its associated modular automorphism. 

Let $L^2(N;\varphi_0)$ be the GNS Hilbert space associated with the reference state. The crossed product $M$ is a von Neumann algebra acting on the extended Hilbert space, $H_{ext} \equiv L^2(N;\varphi_0) \otimes L^2(\mathbb{R})$. In fact, it can be defined as the unique subalgebra of $N \otimes B(L^2(\mathbb{R}))$ which commutes with the operator $h_{\varphi_0} \otimes \mathbb{1}_{L^2(\mathbb{R})} + \mathbb{1}_{L^2(N;\varphi_0)} \otimes \hat{q}$.\footnote{We refer the reader to \cite{Takesaki2,takesaki1973crossed,Klinger:2023auu,AliAhmad:2024wja,AliAhmad:2024vdw} for more detailed introductions to the crossed product and its role in gauge theory and gravity.} Here, $h_{\varphi_0}$ is the (one-sided) modular Hamiltonian of the state $\varphi_0$ which generates the modular automorphism $\sigma^{\varphi_0}$, and $\hat{q}$ is the position operator on $L^2(\mathbb{R})$ which is interpreted as the Hamiltonian of an observer.

Any state $\xi \in L^2(G)$ can be used to define a quantum channel $E_{\xi}: N \otimes B(L^2(\mathbb{R})) \rightarrow N$ as
\beq \label{Observer state extension}
	E_{\xi}(n \otimes \mathcal{O}) = n g_{L^2(G)}(\xi, \mathcal{O} \xi). 
\eeq
The restriction $E_{\xi}\rvert_M: M \rightarrow N$ defines a quantum channel from $M$ to $N$.\footnote{For ease of notation, we will continue to denote this restriction simply by $E_{\xi}$.} The channel $E_{\xi}$ can be used to extend states $\psi_0 \in S(N)$ to states $\psi \equiv \psi_0 \circ E_{\xi} \in S(M)$. As has by now been well documented, the modular crossed product of a type III algebra is semi-finite \cite{takesaki1973crossed}. Thus, there exists a tracial weight\footnote{This choice is generically non-unique resulting in a constant contribution to the entropy. We suppress this term since it is not of central importance to the present discussion.} $\tau \in P(M)$ through which we can define density operators, $\rho_{\psi \mid \tau}$. In \cite{Jensen:2023yxy} such density operators were computed explicitly as
\beq \label{Modular density operator}
	\rho_{\psi \mid \tau} = \bigg( e^{\hat{q}/2} \overline{\xi}(\hat{q} - h_{\varphi_0}) e^{-\hat{p} h_{\varphi_0}}\bigg)^{\dagger} \rho_{\psi_0 \mid \varphi_0} \bigg( e^{\hat{q}/2} \overline{\xi}(\hat{q} - h_{\varphi_0}) e^{-\hat{p} h_{\varphi_0}}\bigg) \equiv \bigg(\rho_{E_{\xi} \mid \tau}^{1/2}\bigg)^{\dagger} \rho_{\psi_0 \mid \varphi_0} \rho_{E_{\xi} \mid \tau}^{1/2}. 
\eeq
Here $\hat{p}$ is the momentum operator on $L^2(\mathbb{R})$. We have also defined $\rho^{1/2}_{E_{\xi} \mid \tau}$ as the (relative) density operator associated with the channel $E_{\xi}$. Eqn.~\eqref{Modular density operator} is another instance of Eqn.~\eqref{Non-Commutative Conditional Factorization}. 

The entropy of a state $\psi$ as defined above was also computed in \cite{Jensen:2023yxy} under the assumption that $[\xi(\hat{q}-h_{\varphi_0}), \rho_{\psi_0 \mid \varphi_0}] \simeq 0$. This corresponds to treating the entanglement between the observer and the original field theory degrees of freedom as negligible. Under this assumption, we find
\beq \label{Entropy in Mod}
    S(\psi) = -D(\psi_0 \parallel \varphi_0) + S(\xi) + \omega^{H_{ext}}_{\xi_{\psi}}(-\hat{h}_{obs}).
\eeq
The first term is the relative entropy between $\psi_0$ and $\varphi_0$ treated as states on the original algebra $N$. The second term is the entropy of the observer's wavefunction, $\xi$. Finally, the third term is the expectation value of the observer's Hamiltonian $\hat{h}_{obs} \equiv \mathbb{1}_{L^2(N;\varphi_0)} \otimes \hat{q}$ in the state $\psi$.   

To relate \eqref{Entropy in Mod} to the familiar form of the generalized entropy we evoke the so-called first law of subregions which tells us that the observer's Hamiltonian is related to the bounding area of the subregion $\Sigma$ by\footnote{We have absorbed $G_N$ into the definition of $A$.}
\beq  \label{First law}
    h_{\varphi_0} + \hat{h}_{obs} = -A/4. 
\eeq
By the Bisignano-Wichmann theorem and its variants, the operator $h_{\varphi_0}$ may be interpreted as the generator of boosts. Recall that the crossed product is the subalgebra of $N \otimes B(L^2(\mathbb{R}))$ which commutes with the operator defined above. We interpret this as a quantum mechanical implementation of the diffeomorphism constraint of quantum gravity applied to boosts. Inserting Eqn.~\eqref{First law} into Eqn.~\eqref{Entropy in Mod} we find
\beq \label{Mod Entropy 2}
    S(\psi) = -D(\psi_0 \parallel \varphi_0) + \omega^{L^2(N;\varphi_0)}_{\xi_{\psi_0}}(h_{\varphi_0}) + S(\xi) + \frac{1}{4}\omega^{H_{ext}}_{\xi_{\varphi}}(A).
\eeq
Although it is strictly speaking not well defined, we can evoke a result from type I algebras which tells us that the relative entropy between two states is given by
\beq \label{Type I result}
    D(\psi_0 \parallel \varphi_0) = \omega^{L^2(N;\varphi_0)}_{\xi_{\psi_0}}(\ln \rho_{\psi_0} - \ln \rho_{\varphi_0}) = -S(\psi_0) - \omega^{L^2(N;\varphi_0)}_{\xi_{\psi_0}}( h_{\varphi_0}). 
\eeq
Here, we have used the heuristic $\rho_{\varphi_0} \sim e^{-h_{\varphi_0}}$. Rearranging this expression we find
\beq
    S(\psi_0) = -D(\psi_0 \parallel \varphi_0) + \omega^{L^2(N;\varphi_0)}_{\xi_{\psi_0}}(h_{\varphi_0}),
\eeq
which allows us to formulate Eqn.~\eqref{Mod Entropy 2} as
\beq \label{Mod Entropy 3}
    S(\psi) = S(\psi_0) + S(\xi) + \frac{1}{4}\omega^{H_{ext}}_{\xi_{\varphi}}(A). 
\eeq

Eqn.~\eqref{Mod Entropy 3} can be read as an instance of Eqn.~\eqref{General Generalized Entropy Formula}. In this physical scenario the operator appearing in the computation of the quantum conditional entropy is truly the physical area. The `entropy' of the state $\psi_0$ on the type III algebra $N$ is ill defined and divergent. This is the same mechanism which we anticipated in Eqn.~\eqref{General Generalized Entropy Formula}, and provides some more context for how to interpret $D(\psi_0 \parallel \tau \circ i)$ in the case that $\tau \circ i$ does not restrict to a faithful, semi-finite, normal trace. Referring to Eqn.~\eqref{Entropy in Mod} we see that the entropy of the state $\psi$ in the semi-finite algebra $M$ \emph{is} finite. To get from Eqn.~\eqref{Entropy in Mod} to Eqn.~\eqref{Mod Entropy 3} we have reconstituted the relative entropy into a divergent entropy by incorporating the expectation value of the one sided modular Hamiltonian as in Eqn.~\eqref{Type I result}. This renders the first term in Eqn.~\eqref{Mod Entropy 3} divergent which forces us to conclude that the area is divergent in a compensating manner.\footnote{Equivalently, $G_N \rightarrow 0$.}

From the point of view of our general framework, the use of the channel in Eqn.~\eqref{Observer state extension} as a state extension and the assumption that the observer is negligibly entangled with the rest of the theory correspond to truncating the quantum conditional entropy at leading order. In our most general result, Eqn.~\eqref{Non-Commutative Conditional Factorization}, the state extension $i_{\varphi}^{\dagger}$ needn't arise from the partial trace of a pure state. Following this approach, we should be capable of constructing more exotic objects than classical-quantum states. The entropy computation in Eqn.~\eqref{Quantification of DPI} also provides some indication on how the entropy of states in the modular crossed product may be enhanced by including higher order contributions rather than assuming that the entanglement between the observer and the field theory degrees of freedom is negligible. Eqn. \eqref{First law} tells us that the observer's Hamiltonian may be interpreted as playing the role of a regulated area operator. When quantum gravity effects are present we should not expect this operator to commute with QFT observables \cite{Ciambelli:2023mir,Ciambelli:2024swv}. 

In future work we hope to consider the problem of directly computing and interpreting the entropy for a more general class of states in the modular crossed product. One of the interesting questions which arises from such a computation is the interpretation of the quantum conditional entropy. For classical quantum states this quantity is directly related to the expectation value of a geometric area operator. Presumably, if we continue to work with classical quantum states but lift the assumption that the observer and the field theory are disentangled, the conditional relative density operator will still be an exponential of the area. Then, the resulting quantum conditional entropy will have an interpretation as the expectation value of the naive area plus a series of corrections that come from the interaction between the observer and the field theory. However, if we work with entirely general states it is not immediately clear that the conditional density operator will continue to be equal to the exponential of the area. We revisit this question in the sections below.

\subsection{Looking Forward}

We conclude this section by exploring two further applications of our framework to holographic subregion duality and quantum black hole interiors. In these discussions, we emphasize how our computational technique allows for existing results, such as the quantum extremal surface prescription and the generalized second law, to be enhanced and understood in more realistic contexts than toy examples. 

\subsubsection{Holographic subregion duality}

Starting from a boundary subregion, $R$, in a holographic CFT we can ask what subregion of the bulk, $r$, corresponds to $R$ under holographic duality. Given any boundary subregion $R$, let $\Gamma(R)$ denote the set of codimension-two surfaces in the bulk which share a boundary with $R$, e.g. $\partial \gamma = \partial R$. We may therefore glue $R$ and $\gamma$ together along their shared boundary and take the resulting manifold to be the boundary of a bulk subregion $r(R,\gamma)$ such that $\partial r(R,\gamma) = R \cup \gamma$. The QES prescription~\cite{Engelhardt:2014gca} proposes that the bulk subregion dual to $R$ is of the form $r(R,\gamma)$ for the surface $\gamma$ which extremizes the generalized entropy:
\beq
	S_{gen}(r(R,\gamma)) \equiv S(r(R,\gamma)) + \frac{A(\gamma)}{4 G_N}.
\eeq
Here, $S(r(R,\gamma))$ is the von Neumann entropy of quantum fields propagating inside the region, and $A(\gamma)$ is the metrical area of the surface $\gamma$. That is
\beq
	r_* = r(R, \gamma_*), \qquad \gamma_* = \text{arg min}_{\gamma \in \Gamma(R)} S_{gen}(r(R,\gamma)). 
\eeq
The resulting bulk subregion, $r_*$, is called the entanglement wedge. 

The QES has provided a robust tool for characterizing holographic duality, but there are some indications that it may be augmented away from the semiclassical limit. Firstly, there are well documented corrections to the generalized entropy contributing at higher order in $\hbar$ and $G_{\rm N}$. These additional terms arise from higher curvature terms of the `area', higher order terms in the matter entropy, and backreaction between the two. Secondly, the current form of the QES implicitly relies upon the existence of a semiclassical bulk dual to define a notion of metrical area and compute the generalized entropy. A more satisfying prescription might define these quantities entirely at the level of the boundary theory. 

The toolkit we have developed in this note seems poised to shed light on both of these problems. To the first point, even in the semiclassical limit our approach is capable of encoding higher order corrections to the generalized entropy. In this regard, even without a more drastic reassessment of the QES, one may be able to track corrections to the entanglement wedge arising from these quantum corrections to the area and entropy. More speculatively, however, our approach provides a way of constructing a generalized entropy which is purely algebraic, with the `area' term acquiring a geometric meaning only upon specifying a particular physical scenario. In turn, we can use our technique to propose a purely algebraic version of the quantum extremal surface prescription. 

Let us outline how this approach may work now. We begin with the algebra of simple operators on a boundary subregion $R$ which we refer to as $S_R$. One may think of $S_R$ as dual to the algebra of the \emph{causal wedge} in the bulk, $CW_R$. The causal wedge is the bulk domain of dependence of the boundary subregion $R$. Bulk observables in $CW_R$ are simple because they can be reconstructed using causality alone without evoking holography. The QES prescription points to an algebra of observables \emph{deeper} in the bulk than the causal wedge to which the full boundary subregion is dual, this is the entanglement wedge $EW_R$. From the perspective of the bulk, the causal wedge is \textit{contained} in the entanglement wedge, $CW_R \subset EW_R$. However, they are both anchored to $R$. If holography respects inclusions, then there must also exist a boundary subalgebra $N_R$ which extends the algebra of simple operators $S_R$, so that $S_R \subset N_{R}$ is dual to the inclusion $CW_R \subset EW_R$ in an appropriate sense. We stress that this inclusion is \textit{not} local since both have support on the same boundary subregion. 

Schematically, we might think of the algebra $N_R$ as being obtained from $S_R$ by adding some collection of non-local operators:
\beq
	N_R \equiv S_R \vee \{\Lambda_i\}_{i \in \mathcal{I}}.
\eeq
More to the point, there can be several different choice of non-local operators which can be appended to the simple algebra. Let's denote by $\Gamma$ the collection of sets of non-local operators which can be chosen to extend the simple algebra.\footnote{In forthcoming work, we provide a rigorous construction of this collection through a generalization of Longo's Q-system construction \cite{ali-ahmad2024_QSystem}.} For each $\gamma = \{\Lambda_i\}_{i \in \mathcal{I}} \subset \Gamma$ we get a different algebra $N_R(\gamma) \equiv S_R \vee \gamma$. Let $i_\gamma: S_R \hookrightarrow N_R(\gamma)$ denote the inclusion of $S_R$ into the extended algebra for a choice of non-local operators, $\gamma$. Provided $N_R(\gamma)$ is a semi-finite algebra, we can use our non-commutative Bayes' theorem to factorize the entropy of a state $\varphi \in S(N_R(A))$ into the sum of contributions:
\beq \label{Algebraic QES Entropy}
	S(\varphi; \gamma) = -D(\varphi \parallel \tau \circ i_A) + \varphi_0(A_{i_{\gamma,\varphi}^{\dagger}}). 
\eeq
We think of Eqn.~\eqref{Algebraic QES Entropy} as the generalized entropy associated with the extension $\gamma$. It consists of the entropy contributed by the simple observables plus an `area' term which encodes the contribution of the non-local operators in $\gamma$.

Mimicking the structure of the QES, we can define the algebra which is dual to the entanglement wedge in the bulk as the extension of the simple algebra by non-local operators which extremizes the quantity Eqn.~\eqref{Algebraic QES Entropy}:
\beq \label{Algebraic QES}
	EW_R \simeq N_R(\gamma_*), \qquad \gamma_* \equiv \text{arg min}_{\gamma \in \Gamma} \bigg( \text{min}_{\varphi \in S(N_R(\gamma))} S(\varphi;\gamma) \bigg). 
\eeq  
Here we have also included an extremization over states in each fixed candidate algebra. Presumably, a further restriction must be placed on the set of extensions $\gamma$ and states $\varphi$ which are allowed in the above extremization to ensure that we do not just revert to the non-extended algebra as the entropy minimizer. This is akin to the requirement that the extremal surface share a boundary with the boundary subregion in the semiclassical form of the QES. Eqn.~\eqref{Algebraic QES} provides a fully algebraic ``QES prescription". In future work, we plan to investigate whether it can be used to reproduce semiclassical results in the appropriate limits. If so, one can argue that this proposal provides a fully algebraic prediction of the emergence of the entanglement wedge from the boundary.  It can also realize different reconstruction schemes by extremizing different entropic quantities.

\subsubsection{Quantum black holes}
A source of many puzzles in quantum gravity is the nature of black holes. The introduction of quantum contributions to the geometric description of black holes leads to many obstacles that defeat our classical intuition. For a nice review, see Ref.~\cite{Harlow:2014yka}. In this section, we would like to briefly outline how our main result sheds light on two important and related aspects of quantum black holes: (1) the generalized second law and (2) black hole evaporation. 

Wall's seminal proof of the generalized second law depends intimately upon the data processing inequality \cite{Wall:2011hj}. One of the major contributions of this work is an exact quantification of the information gap appearing in the data processing inequality \eqref{Quantification of DPI}. In this way, we can enhance some of the interpretation of the generalized second law by providing a quantification of the information gap which appears in this context. Wall's original argument relied upon the existence of a renormalization scheme that regulates the ultraviolet divergences of entropies he computed. At the time of his derivation this was an assumption. We now understand that the crossed product provides precisely such a universal renormalization scheme for the generalized entropy~\cite{Chandrasekaran:2022eqq, AliAhmad:2023etg,Faulkner:2024gst}. A recent paper by Faulkner and Speranza~\cite{Faulkner:2024gst} explicitly reformulates Wall's argument in the language of the crossed product. 

Let us provide a simplified version of how this argument works, we refer the reader to \cite{Faulkner:2024gst} for the detailed version. We begin by defining algebras of observables $M_{\Lambda}$ and $M_{\Lambda'}$ for a semiclassical gravity theory on different cuts dictated by the null parameter on the horizon of a black hole, $\lambda = \Lambda, \Lambda'$. Based on the above discussion, we should view the algebras $M_{\Lambda}$ as modular crossed products of effective field theory algebras $N_{\Lambda}$ and $N_{\Lambda'}$. Without loss of generality, we assume that $\Lambda' > \Lambda$ which implies that $M_{\Lambda} \subset M_{\Lambda'}$. In analogy with the computation we have reviewed in Section \ref{sec: Modular}, the von Neumann entropy in the crossed product is a sum of a relative entropy term and a term that depends on the fluctuations of the asymptotic charges, in this case the mass of the black hole. Thus, the inclusion structure $M_{\Lambda} \subset M_{\Lambda'}$ essentially proves the generalized second law through the monotonicity of the entropy, which itself is a consequence of the data processing inequality. 

Let $i: M_{\Lambda} \hookrightarrow M_{\Lambda'}$ be the inclusion of horizon algebras, and let us assume that the trace $\tau \in P(M_{\Lambda}')$ restricts to a trace $\tau \circ i \in P(M_{\Lambda})$. Then, given any state $\varphi \in S(M_{\Lambda'})$ which restricts to $\varphi \circ i \in S(M_{\Lambda})$ we can compute the difference in their entropies by invoking Eqn.~\eqref{Quantification of DPI}:
\beq
	D(\varphi \parallel \tau) - D(\varphi \circ i \parallel \tau \circ i) = \sum_{n = 0}^{\infty} \frac{c_n}{n!} \omega_{\xi_{\psi}}^{L^2(M;\varphi)} \bigg(\text{ad}_{\ln \rho_{\alpha_{\psi}^{\dagger} \mid \alpha,\varphi}}^n(\ln \rho_{\psi \circ \alpha \mid \varphi \circ \alpha}) \bigg). 
\eeq
In future work, we hope to return to a concrete computation of the terms appearing on the right hand side of the above equation. For example, in the case of a Rindler horizon, we expect this computation to reproduce higher order terms in the semiclassical expansion in $\hbar G_{\rm N}$.

Moving on to black hole evaporation, many recent works have proposed algebraic approaches to this problem~\cite{gomez2024algebraicpagecurve,Gomez:2024cjm,vanderHeijden:2024tdk}. Unfortunately, all of these rely upon the notion of the index as a measure for how radiation sits inside of the effective description of the black hole. The existence of a finite index is directly related to the existence of what we have referred to as strict conditional expectations, which implement an exact recovery of black hole degrees of freedom from the radiation. As we have stressed, the existence of strict conditional expectations relies upon strong assumptions that aren't typically met in realistic models of black holes. By contrast, our main result remains valid even in this regime by passing from strict to generalized conditional expectations. Our approach may therefore provide a quantification of the information deficit between the black hole and its radiation even when these degrees of freedom are described by a total algebra which does \textit{not} factorize, improving upon what is typically assumed in finite models. In this context, the role of the index is replaced by the quantum conditional entropy which provides a more complicated structure compatible with approximate recovery.

\section{Discussion} \label{sec: Discussion}

The overarching aim of this note has been to uncover quantum analogs for classical notions from conditional probability theory. The three main concepts that we sought to adapt were
\begin{enumerate}
	\item The factorization of a classical (joint) probability distribution into the product of a marginal and a conditional piece:
	\beq \label{Factorization Discussion}
		p(x_1,x_2) = p_1(x_1) p_{2 \mid 1}(x_2 \mid x_1).
	\eeq
	\item A quantification of the classical data processing inequality in terms of the relative conditional entropy:
	\beq \label{Classical DPI Discussion}
		D(p \parallel q) - D(p_1 \parallel q_1) = \mathbb{E}_{p_1}\bigg(D(p_{2 \mid 1} \parallel q_{2 \mid 1})\rvert_{X_1}\bigg). 
	\eeq
 \item The factorization of the Shannon entropy into a sum of a marginal and a conditional contribution:
	\beq \label{Entropy Factorization Discussion}
		S(p) = S(p_1) + \mathbb{E}_{p_1}\bigg(S(p_{2 \mid 1})\rvert_{X_1}\bigg).
	\eeq	
\end{enumerate} 
The three main results of our work accomplish this goal: 
\begin{enumerate}
    \item Given a quantum channel $\alpha: N \rightarrow M$ and a pair of faithful, semi-finite, normal states $\varphi,\psi \in S(M)$ whose compositions $\varphi \circ \alpha, \psi \circ \alpha \in S(N)$ are also faithful, semi-finite, and normal, the relative density operator $\rho_{\psi \mid \varphi}$ can be expressed as a non-commutative factorization\footnote{This is provided the Petz dual map $\alpha_{\psi}^{\dagger}: M \rightarrow N$ is differentiable with respect to $\alpha_{\varphi}^{\dagger}: M \rightarrow N$.}
\beq \label{NC Factorization Discussion}
	\rho_{\psi \mid \varphi} = \bigg(\rho_{\alpha_{\psi}^{\dagger} \mid \alpha, \varphi}^{1/2}\bigg)^{\dagger} \rho_{\psi \circ \alpha \mid \varphi \circ \alpha} \; \rho_{\alpha_{\psi}^{\dagger} \mid \alpha, \varphi}^{1/2}. 
\eeq
Here, $\rho_{\psi \circ \alpha \mid \varphi \circ \alpha}^{1/2}$ is the `marginal' relative density operator between $\psi$ and $\varphi$ after the application of the channel. We refer to $\rho_{\alpha_{\psi}^{\dagger} \mid \alpha,\varphi}^{1/2}$ as the `conditional' relative density operator. It is constructed from the spatial derivative of $\alpha_{\psi}^{\dagger}$ by $\alpha_{\varphi}^{\dagger}$, and spatial implementations of the channels $\alpha$ and $\alpha_{\varphi}^{\dagger}$ \eqref{Density factorization for channels}. Eqn.~\eqref{NC Factorization Discussion} is a non-commutative analog of Eqn.~\eqref{Factorization Discussion}, which reduces to the classical Bayes' law when the underlying algebras are Abelian. 
\item 
Using Eqn.~\eqref{NC Factorization Discussion}, one can define a notion of  quantum conditional entropy which precisely quantifies the information gap appearing in the data processing inequality for arbitrary quantum channels:
\beq \label{DPI Discussion}
    D(\psi \parallel \varphi) - D(\psi \circ \alpha \parallel \varphi \circ \alpha) = \omega_{\xi_{\psi}}^{L^2(M;\varphi)}\bigg(\ln \rho_{\alpha_{\psi}^{\dagger} \mid \alpha,\varphi}\bigg) + \sum_{n = 1}^{\infty} \frac{c_n}{n!} \omega_{\xi_{\psi}}^{L^2(M;\varphi)} \bigg(\text{ad}_{\ln \rho_{\alpha_{\psi}^{\dagger} \mid \alpha,\varphi}}^n(\ln \rho_{\psi \circ \alpha \mid \varphi \circ \alpha}) \bigg) 
\eeq
This is the quantum analog of Eqn.~\eqref{Classical DPI Discussion}. 

\item A special case of \eqref{DPI Discussion} factorizes the entropy of any state $\varphi \in S(M)$ in a semi-finite algebra $M$ as a sum of terms
\beq \label{Gen Ent Discussion}
    S_{\tau}(\varphi) = -D(\varphi \parallel \tau \circ i) + \varphi_0( A_{i_{\varphi}^{\dagger}}). 
\eeq
Eqn.~\eqref{Gen Ent Discussion} is the quantum analog of \eqref{Entropy Factorization Discussion}. 

\end{enumerate}

A crucial insight underlying these results is the observation that generalized conditional expectations (or more broadly Petz dual maps) provide a mechanism for factorizing states on von Neumann algebras as the composition of a marginal state and a quantum channel \emph{without} requiring a strict factorization of the underlying algebra. This is in direct contrast to the standard notion of conditional expectation, whose homogeneity property predicates a high degree of factorization. Such a factorization fundamentally suppresses quantum entanglement within the algebra. This forces the associated notion of quantum conditional entropy to be expressed as the expectation value of an operator affiliated with an algebraic center, which is an essentially classical (commutative) object. The general notion of quantum conditional entropy we've introduced \eqref{DPI Discussion} does not suffer from this issue and can be formulated as the expectation value of an operator which is affiliated with the algebras in question, only becoming central in the event that the channel of interest is a strict conditional expectation. 

It is natural to interpret the transition from strict conditional expectations to generalized conditional expectations in the terminology of approximate quantum error correction. As a unital and homogeneous map, the standard conditional expectation plays the role of an effective inverse for its associated inclusion. More specifically, given an inclusion of von Neumann algebras $i: N \hookrightarrow M$ admitting a conditional expectation $E: M \rightarrow N$, operators $n \in N$ are protected under the action of $E$ when viewed as a quantum channel so that $E \circ i(n) = n$. This fact is intimately tied to the factorization of the algebra implied by the homogeneity of the conditional expectation. On the other hand, given a generalized conditional expectation constructed as the Petz dual of the inclusion with respect to a state $\varphi \in S(M)$, $i_{\varphi}^{\dagger}: M \rightarrow N$, we have that $i_{\varphi}^{\dagger} \circ i(n) = n$ only for a subset $N_{i,\varphi} \subset N$ in which the factorization property of the conditional expectation is met. We interpret this to mean that under the existence of a generalized conditional expectation, only \emph{some} of the operators in $N$ are perfectly recoverable. Equivalently, the larger algebra $M$ admits only a partial factorization. 

The quantum conditional relative entropy between states $\varphi, \psi \in S(M)$ with respect to a channel $\alpha: N \rightarrow M$ measures the obstruction to the channel's sufficiency. Recall that a channel is sufficient with respect to a pair of states if and only if the states are as distinguishable after applying the channel as they were before. In the case that $\alpha: N \hookrightarrow M$ is an inclusion, we take this to mean that the subalgebra $N$ is sufficient for distinguishing between $\varphi$ and $\psi$. In other words, one can reliably judge $\varphi$ and $\psi$ as distinct states using only measurements made with respect to observables in the algebra $N$. Being able to precisely quantify the difference between the distinguishability of states before and after the application of a quantum channel provides an invaluable tool for understanding state based quantum error correction and quantum statistical inference. 

Shifting our attention to Eqn.~\eqref{Gen Ent Discussion}, an important observation deserves to be emphasized. The first term therein can be interpreted as a notion of entropy for the state $\varphi$ when restricted to a subalgebra $i: N \hookrightarrow M$. This `entropy' can be reduced to a standard algebraic entropy with respect to a tracial weight whenever $N$ is semi-finite, but will be divergent in the event that $N$ is type III. The second term in Eqn.~\eqref{Gen Ent Discussion} is interpreted as the quantum conditional entropy of the state $\varphi$. It is expressed as the expectation value of an operator $A_{i_{\varphi}^{\dagger}}$. In the case that $N$ is semi-finite, one should expect this operator to live explicitly inside of $N$. On the other hand, if $N$ is type III, the operator $A_{i_{\varphi}^{\dagger}}$ becomes unbounded and is merely affiliated with $N$. We arrive at this conclusion because the left hand side of Eqn.~\eqref{Gen Ent Discussion} is finite up to a possibly divergent state independent constant, but, as discussed above, the first term on the right hand side is automatically divergent if $N$ is not semi-finite. Thus, it must be the case that the divergence of this term is canceled by a compensating divergence in the quantum conditional entropy.

In physical contexts, Eqn.~\eqref{Gen Ent Discussion} provides an interesting proposal for the generalized entropy which reconciles an apparent tension. In particular, it was originally argued that the generalized entropy should emerge from a factorization of the entropy as informed by the existence of a strict conditional expectation \cite{faulkner2020holographic}. It has since been shown that the generalized entropy is realized naturally in gravitational contexts by evoking properties of the modular crossed product such as its semi-finiteness and its interpretation as a gauge invariant algebra. The modular crossed product is an extension of a QFT subregion algebra, and thus admits an inclusion $i: N \hookrightarrow M \equiv N \times_{\sigma^{\varphi}} \mathbb{R}$. However, by standard arguments $N$ is a type III$_1$ algebra, while $M$ is semi-finite. A theorem by Stormer \cite{stormer1997conditional} states that there cannot be a faithful conditional expectation mapping from $M$ to $N$. Thus, the generalized entropy formula emerging in the context of the modular crossed product cannot be traced back to a conditional expectation as in \cite{faulkner2020holographic}. Fortunately, we can still construct a generalized conditional expectation between $M$ and $N$, and in doing so reproduce the entropy factorization of the modular crossed product from our more broad point of view. As we have discussed, this has the effect of promoting the area from a central to a generally non-commuting operator. 

In future work we hope to explore the physical interpretation of our area operator further. The result we have derived is valid for arbitrary von Neumann algebras, and thus one may hypothesize that it should be capable of encoding the full score of quantum corrections to the generalized entropy. Away from the semiclassical limit, it is no longer reasonable to regard $A_{i_{\varphi}^{\dagger}}$ as coinciding with the metrical area of an extremal surface. It is tempting, therefore, to instead regard $A_{i_{\varphi}^{\dagger}}$ as \emph{defining} the `area' in a fully quantum manner. This provides a very satisfying avatar for the notion of geometry as emerging entirely from entanglement.

\appendix
\renewcommand{\theequation}{\thesection.\arabic{equation}}
\setcounter{equation}{0}

\section*{Acknowledgments}

S.A.A. would like to thank Ahmed Almheiri, Ro Jefferson, and Yifan Wang for insightful conversations. M.S.K. would like to thank Thomas Faulkner, Samuel Goldman, Jonah Kudler-Flam, Robert Leigh, Juan Maldacena, and Edward Witten for helpful discussions and comments.

\providecommand{\href}[2]{#2}\begingroup\raggedright\endgroup

\end{document}

%% file: packages.tex
\usepackage{amsmath,amssymb,amsfonts,bbm,bm}
\usepackage{slashed}
\usepackage{braket}
\usepackage{tocloft} 

\usepackage[nosort]{cite} 
\usepackage{graphicx,color}
\usepackage[colorlinks=true, linkcolor=blue, citecolor=blue, linktoc=all]{hyperref}

\usepackage[usenames,dvipsnames]{xcolor}

\usepackage{tikz-cd}
\usepackage{pict2e}
\usepackage{physics}
\usepackage{comment} 
\usepackage[makeroom]{cancel}

%% file: standard.tex
\newcommand{\beq}{\begin{eqnarray}}
\newcommand{\eeq}{\end{eqnarray}}
\newcommand{\beqn}{\begin{eqnarray}}
\newcommand{\eeqn}{\end{eqnarray}}
\newcommand{\bea}{\begin{eqnarray}}
\newcommand{\eea}{\end{eqnarray}}
\newcommand{\be}{\begin{equation}}
\newcommand{\ee}{\end{equation}}

\usepackage[normalem]{ulem}

\newcommand{\uiuc}[1]{
	\centerline{
		\begin{minipage}[c]{0.7\textwidth}
			\begin{center}
			${}^{#1}$ Illinois Center for Advanced Studies of the Universe \& Department of Physics,\\ 
			University of Illinois, 1110 West Green St., Urbana IL 61801, U.S.A.
			\end{center}
		\end{minipage}
		}
	}
   \newcommand{\nyu}[1]{
	\centerline{
		\begin{minipage}[c]{0.7\textwidth}
			\begin{center}
			${}^{#1}$ Center for Cosmology and Particle Physics, New York University, New York, NY 10003, USA
			\end{center}
		\end{minipage}
		}
	}

%% file: AlgMacros.tex
\usepackage{mathrsfs,mathtools}
\renewcommand\mathbb[1]{\mathbbm{#1}}

\makeatletter
\DeclareRobustCommand{\loplus}{\mathbin{\mathpalette\dog@lsemi{+}}}
\DeclareRobustCommand{\lotimes}{\mathbin{\mathpalette\dog@lsemi{\times}}}
\DeclareRobustCommand{\roplus}{\mathbin{\mathpalette\dog@rsemi{+}}}
\DeclareRobustCommand{\rotimes}{\mathbin{\mathpalette\dog@rsemi{\times}}}

\newcommand{\dog@rsemi}[2]{\dog@semi{#1}{#2}{-90,90}}
\newcommand{\dog@lsemi}[2]{\dog@semi{#1}{#2}{270,90}}
\newcommand{\dog@semi}[3]{%
  \begingroup
  \sbox\z@{$\m@th#1#2$}%
  \setlength{\unitlength}{\dimexpr\ht\z@+\dp\z@\relax}%
  \makebox[\wd\z@]{\raisebox{-\dp\z@}{%
    \begin{picture}(1,1)
    \linethickness{\variable@rule{#1}}
    \roundcap
    \put(0.5,0.5){\makebox(0,0){\raisebox{\dp\z@}{$\m@th#1#2$}}}
    \put(0.5,0.5){\arc[#3]{0.5}}
    \end{picture}%
  }}%
  \endgroup
}
\newcommand{\variable@rule}[1]{%
  \fontdimen8  
  \ifx#1\displaystyle\textfont3\else
    \ifx#1\textstyle\textfont3\else
      \ifx#1\scriptstyle\scriptfont3\else
        \scriptscriptfont3\relax
  \fi\fi\fi
}
\DeclareRobustCommand{\loplus}{\mathbin{\mathpalette\dog@lsemi{+}}}